\documentclass[journal]{IEEEtran}
\newcommand{\ignore}[1]{}
\usepackage[T1]{fontenc}
\usepackage{amsmath,amssymb,amsfonts}
\usepackage{graphicx}
\usepackage{booktabs}
\usepackage{array}
\usepackage{multirow}
\usepackage{url}
\usepackage[table]{xcolor}
\usepackage{algorithmic}
\usepackage{textcomp}
\usepackage{float}
\usepackage{pifont}
\usepackage{adjustbox}
\usepackage{hhline}
\usepackage{balance}
\usepackage{mathtools}

\usepackage{courier}
\usepackage{ragged2e}
\usepackage{tikz}
\usetikzlibrary{shapes,arrows,positioning}
\usepackage{orcidlink}

\usepackage{hyperref}
\hypersetup{
  colorlinks,
  allcolors=blue,
  citebordercolor=blue,
  citecolor=blue,
  linkcolor=blue,
  urlcolor=blue}

\definecolor{headerblue}{RGB}{31,73,125}
\definecolor{rowgray}{RGB}{242,242,242}

\usepackage{setspace}
\begin{document}
\bstctlcite{IEEEexample:BSTcontrol}
\title{Hardware-Enabled Fuzzy Inference:\\ Architectures, Platforms, and Emerging Trends}
\author{
Amir Hossein Jalilvand~\orcidlink{0000-0002-7641-6606},
Parsa Hassani Shariat Panahi~\orcidlink{0009-0005-2912-3754},
and M. Hassan Najafi~\orcidlink{0000-0002-4655-6229} 
\vspace{-1em}
\thanks{Amir Hossein Jalilvand and Parsa Hassani Shariat Panahi are with the School of Computer Engineering, Iran University of Science and Technology, Tehran, Iran. M. Hassan Najafi is with the Electrical, Computer, and Systems Engineering Department, Case Western Reserve University, OH, USA. 
Corresponding Author:~M.~Hassan Najafi~najafi@case.edu
}}

\maketitle
\begin{abstract}
Fuzzy logic systems are widely used for intelligent decision-making under uncertainty, offering interpretability and robustness across diverse applications. However, the growing demand for real-time edge intelligence has exposed the limitations of software-based fuzzy inference: unpredictable latency, excessive power consumption, and inefficient resource utilization. This has motivated extensive research into hardware acceleration, spanning platforms from custom analog circuits and digital ASICs to reconfigurable FPGAs and ultra-low-power microcontrollers. This survey presents a 
comprehensive, platform-centric review of hardware fuzzy systems, systematically organizing the literature into three principal categories: FPGA-based implementations, ASIC and custom VLSI realizations, and embedded, IoT, and TinyML platforms. For each category, we analyze architectural organization, resource mapping strategies, implementation trade-offs, and key design challenges.

Our cross-platform comparative analysis reveals that no single platform dominates across all metrics. FPGAs offer flexibility and rapid prototyping, ASICs deliver peak performance and energy efficiency, while embedded and TinyML systems balance low power and cost for edge deployment. Despite significant progress, critical research gaps persist: the absence of standardized benchmarks, limited scalability of rule bases, insufficient design automation, and limited support for online learning and emerging memory technologies. We outline future directions, including in-memory fuzzy computing with memristive crossbars, integration with TinyML ecosystems, explainable hardware AI, and open-source design automation. This survey serves as a reference for researchers and practitioners working on hardware-enabled fuzzy intelligence.
\end{abstract}


\vspace{-1em}
\section{Introduction}\label{sec:introduction}

\IEEEPARstart{F}{uzzy} logic systems (FLS) are widely used for intelligent decision-making and control under uncertainty, offering a mathematically rigorous framework for modeling human reasoning through linguistic variables, membership functions, and rule-based inference \cite{zadeh1994softcomputing, lee1990fuzzy, lee1990fuzzy2, driankov1993fuzzy, nethaji2024performance, lu2024fuzzy, alateeq2024logic, evolving2025fuzzycontrol, infus2024proceedings, akbari2024fmcdm}. Unlike conventional binary logic, fuzzy logic accommodates partial truth and graded membership, making it particularly effective for nonlinear, poorly modeled, or imprecise systems \cite{de2024fpga}. Despite advances in deep learning, fuzzy systems remain highly relevant for applications demanding \textit{interpretability}, \textit{low-latency}, and \textit{reliable operation in resource-constrained environments} \cite{lu2024fuzzy,neelu2025reconfigured, tran2026high}. The transparency of fuzzy inference, where each decision traces back to explicit rules, offers a distinct advantage over black-box neural models, especially in safety-critical domains.

While software implementations are convenient for prototyping, they often prove inadequate for real-time embedded systems due to non-deterministic latency and excessive power consumption \cite{nethaji2024performance}. These limitations have motivated over three decades of research into the \textit{hardware acceleration} of fuzzy systems, targeting platforms from custom analog circuits and digital ASICs to reconfigurable FPGAs and ultra-low-power {microcontrollers (MCUs)} \cite{talaska2019parallel}. The recent growth of edge computing, {the Internet of Things (IoT)}, and TinyML has further amplified interest in compact and energy-efficient fuzzy hardware  \cite{fister2025control, jeong2023design, warden2020tinyml}. Contemporary research has demonstrated fuzzy inference engines mapped to MCUs with sub-kilobyte memory, FPGA-based {systems-on-chips (SoCs)}, and analog circuits achieving sub-microwatt power ~\cite{odry2021stochastic, jalilvand2020fuzzy}. These advances position fuzzy hardware as a practical alternative to neural accelerators for edge intelligence. 
{The case for dedicated fuzzy hardware rests on workload requirements rather than on execution speed alone. Three classes of applications justify hardware acceleration. The first is closed-loop control with sub-microsecond deadlines, such as motor drives and power converters, where software running on a general-purpose processor cannot satisfy the sampling period and jitter constraints of the control loop ~\cite{swami2022fpga}. The second is high-throughput inference at the sensor edge, where the sampling rate exceeds the rate at which a sequential processor can fuzzify, evaluate, and defuzzify the incoming stream, so that the resulting backlog violates the real-time constraint ~\cite{srivastava2026reconfigurable, lin2025implementation}. The third is ultra-low-power operation on battery-constrained nodes, where the microcontroller must remain in sleep mode for most of its lifetime and the energy cost of waking, executing, and returning to sleep dominates the design ~\cite{neelu2025reconfigured, merrikh2014memristive, li2025memristive, ahmad2025memristive}. In each case, hardware is justified not by faster execution in isolation, but by the inability of a software implementation to meet a specific latency, throughput, or energy constraint at acceptable cost and accuracy. The remainder of this survey assesses, for each platform, whether and under what conditions hardware resolves these constraints.}

\begin{figure*}[]
\vspace{1em}
\centering
\includegraphics[width=1.8\columnwidth,trim=0.0cm 0cm 0.0cm 0.0cm,clip]{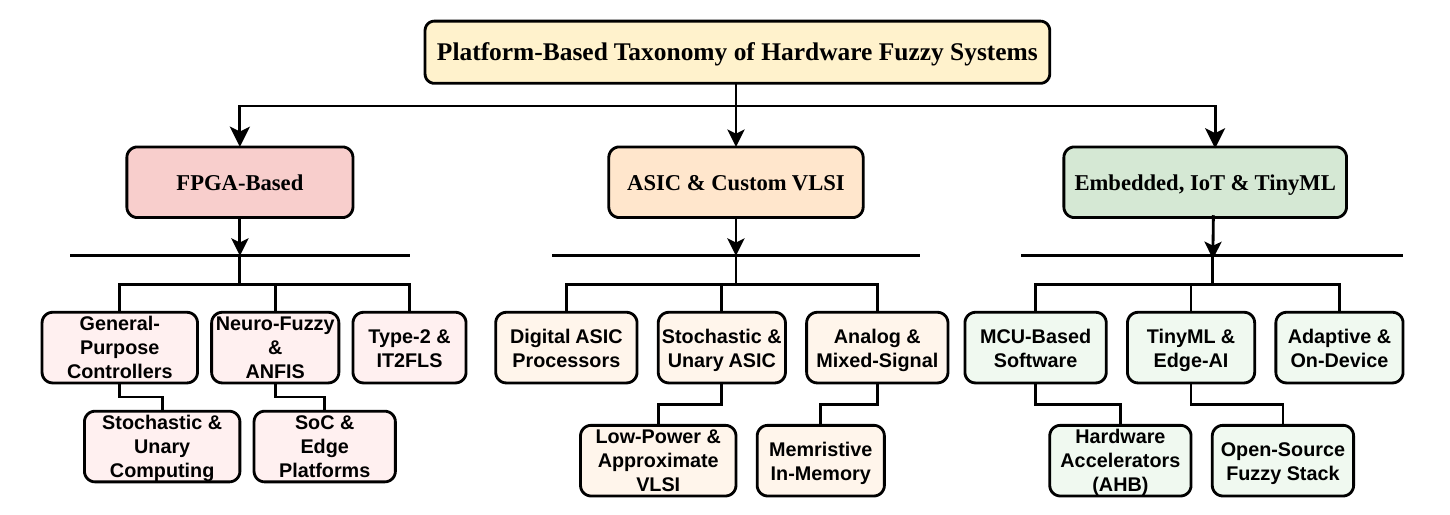}
\vspace{-1em}
\caption{Platform-based taxonomy of hardware fuzzy systems, organizing the literature into three principal categories: FPGA-based, ASIC-based, and embedded/TinyML implementations.}
\label{fig:taxonomy}
\vspace{-1.5em}
\end{figure*}

Existing surveys each answer a narrower question than the one a system
designer faces. Sulaiman \textit{et al.}~\cite{sulaiman2009fpga} and
de Lima and Oliveira~\cite{de2024fpga} ask which applications have been
built on a given platform, and review FPGA-based fuzzy controllers for
general control and for MPPT in photovoltaic panels, respectively.
Murshid \textit{et al.}~\cite{murshid2011vlsi} and
Afzal~\cite{afzal2017hardware} ask which circuit techniques have been
used inside a given component, and review VLSI architectures for
fuzzifiers, inference engines, and defuzzifiers. Valdez
\textit{et al.}~\cite{valdez2023fuzzy} address a tool question: how a
designer can implement a FIS on an FPGA through a graphical interface
without re-synthesizing the design when the rule base changes. None of
these surveys addresses the choice made at the start of a project:
given a latency, power, memory, and cost budget, which platform should
be selected, and what trade-offs does that selection impose? As
Table~\ref{tab:survey_comparison} shows, to the best of our knowledge,
no prior survey covers microcontroller and TinyML platforms, none
compares FPGA, ASIC, and microcontroller implementations within one
framework, and none identifies research gaps across the three platform
classes.

\ignore{
Existing surveys have largely followed
application-driven, component-oriented, or tool-oriented approaches. 
Sulaiman \textit{et al.}~\cite{sulaiman2009fpga} reviewed FPGA-based fuzzy
logic controllers for control applications, focusing on implementation
techniques and experimental results. Murshid \textit{et al.}~\cite{murshid2011vlsi}
and Afzal~\cite{afzal2017hardware} provided component-oriented reviews of
VLSI architectures for fuzzifiers, defuzzifiers, and inference engines.
Valdez \textit{et al.}~\cite{valdez2023fuzzy} presented a tool-oriented
approach for facilitating FIS implementation on FPGAs through a graphical
interface. More recently, de Lima and Oliveira~\cite{de2024fpga}
comprehensively surveyed FPGA-based fuzzy controllers for MPPT in
photovoltaic panels, providing valuable domain-specific insights. As
summarized in Table~\ref{tab:survey_comparison}, none of these surveys
adopts a platform-centric perspective that systematically compares FPGA,
ASIC, and microcontroller/TinyML implementations, and none reports a
cross-platform quantitative comparison across these three platform classes.
}

This survey addresses that question with a platform-centric taxonomy
that organizes the literature into three categories, as illustrated in
Fig.~\ref{fig:taxonomy}: (1) FPGA-based implementations, (2) ASIC and
custom VLSI realizations, and (3) embedded, IoT, and TinyML systems.
For each category, the survey examines architectural organization,
resource mapping, implementation trade-offs, and key challenges. The
platform-centric view makes explicit how the architectural choices at
each pipeline stage determine whether a system meets its speed, power,
area, and flexibility targets.

\ignore{
This survey addresses this gap with a platform-centric taxonomy organizing
the literature into three principal categories, as illustrated in
Fig.~\ref{fig:taxonomy}: (1) FPGA-based implementations, (2) ASIC and custom
VLSI realizations, and (3) embedded, IoT, and TinyML systems. For each category, the survey examines architectural organization, resource mapping, implementation trade-offs, and key challenges. The comparison spans all three platform
classes and, to the best of our knowledge, no prior survey provides such a
comparison in a unified framework. Architectural choices in each stage
determine whether a system meets its speed, power, area, and flexibility
targets, and the platform-centric view makes those choices explicit.
}

The contributions of this survey are fourfold:
\begin{itemize}
\item A comprehensive review emphasizing recent advances in FPGA acceleration, stochastic computing, and TinyML integration.
\item A novel platform-based taxonomy enabling systematic categorization and comparison of fuzzy hardware implementations across diverse technology platforms.
\item A cross-platform comparative analysis identifying the strengths, weaknesses, and application suitability of each hardware category in terms of speed, power, flexibility, and development cost.
\item A critical assessment of research gaps and open challenges, including the lack of standardized benchmarks, limited scalability, insufficient automation for adaptive hardware, and limited use of emerging memory technologies.
\end{itemize}

\begin{table}[]
\centering
\caption{{Comparison of existing surveys on hardware  implementation of fuzzy systems.}}
\label{tab:survey_comparison}
\renewcommand{\arraystretch}{1.3}
\setlength{\tabcolsep}{4pt}
\footnotesize
\begin{tabular}{@{} 
    >{\raggedright\arraybackslash}m{2cm}
    >{\centering\arraybackslash}m{0.6cm}
    >{\raggedright\arraybackslash}m{2cm}
    >{\centering\arraybackslash}m{0.6cm}
    >{\centering\arraybackslash}m{0.7cm}
    >{\centering\arraybackslash}m{0.6cm}
    >{\centering\arraybackslash}m{0.6cm}
    @{}}
\toprule
\rowcolor{headerblue}
{\color{white}\textbf{Survey}} &
{\color{white}\textbf{Year}} &
{\color{white}\textbf{Focus}} &
{\color{white}\textbf{FPGA}} &
{\color{white}\textbf{ASIC}} &
{\color{white}\textbf{MCU}} &
{\color{white}\textbf{Gaps}} \\
\midrule
Sulaiman \textit{et al.}~\cite{sulaiman2009fpga} & 2009 & FPGA-based FLC design & \checkmark & -- & -- & -- \\
Murshid \textit{et al.}~\cite{murshid2011vlsi} & 2011 & VLSI architectures & \checkmark & \checkmark & -- & -- \\
Afzal~\cite{afzal2017hardware} & 2017 & VLSI fuzzy processors & \checkmark & \checkmark & -- & -- \\
Valdez \textit{et al.}~\cite{valdez2023fuzzy} & 2023 & Tool (GUI) for FPGA FIS & \checkmark & -- & -- & -- \\
de Lima \& Oliveira~\cite{de2024fpga} & 2024 & FPGA-based FLC for MPPT & \checkmark & -- & -- & -- \\
\textbf{This survey} &  & \textbf{Platform-centric taxonomy} & \checkmark & \checkmark & \checkmark & \checkmark \\
\bottomrule
\end{tabular}
\vspace{-0.75em}
\end{table}


The remainder of this paper is organized as follows.
{Section~\ref{sec:methodology} describes the review methodology,
including the databases searched, the inclusion and exclusion criteria,
and the screening procedure.} Section~\ref{sec:background} provides the
theoretical background on fuzzy inference systems with emphasis on
hardware-relevant aspects. Section~\ref{sec:taxonomy} presents the
platform-based taxonomy in detail.
Section~\ref{sec:cross-platform} presents a cross-platform
comparison of the three platform classes, and
Section~\ref{sec:discussion} identifies major research gaps. Finally, Section~\ref{sec:conclusion} concludes the paper.

\section{Review Methodology}
\label{sec:methodology}

The literature reviewed in this survey was identified, selected, and
coded following the PRISMA 2020 reporting guideline for systematic
reviews~\cite{page2021prisma}. The guideline is used here as a
reporting framework rather than as a meta-analytic protocol: the
quantitative synthesis items of PRISMA that presuppose homogeneous
outcome measures (e.g., heterogeneity testing, publication-bias
assessment) are not applicable to a heterogeneous hardware corpus, and
they are therefore omitted. The information sources, the full search
strategy, the eligibility criteria, the screening procedure, the
coding scheme, and the resulting study-selection flow are reported
below so that the review can be reproduced and audited.
Table~\ref{tab:prisma} reports the PRISMA 2020 flow in tabular form. 

\begin{table}[t]
\vspace{1em}
\centering
\caption{{PRISMA 2020 flow of the study-selection process, reported
in tabular form. Record counts are those of the search executed in
September 2026 across IEEE Xplore, the ACM Digital Library, and
ScienceDirect.}}
\label{tab:prisma}
\renewcommand{\arraystretch}{1.35}
\setlength{\tabcolsep}{4pt}
\footnotesize
\vspace{-0.5em}
\begin{tabular}{@{}>{\centering\arraybackslash}m{1.5cm}
                  >{\raggedright\arraybackslash}m{2.8cm}
                  >{\raggedright\arraybackslash}m{4cm}@{}}
\toprule
\rowcolor{headerblue}
{\color{white}\textbf{PRISMA stage}} &
{\color{white}\textbf{Records retained}} &
{\color{white}\textbf{Records removed or added}} \\
\midrule
{\textbf{Identification}} &
{Records identified from databases ($n{=}412$): IEEE Xplore
$198$, ACM Digital Library $121$, ScienceDirect $93$.} &
{Duplicates removed ($n{=}74$). Records remaining after duplicate
removal: $n{=}338$.} \\
\rowcolor{rowgray}
{\textbf{Screening}} &
{Records screened on title and abstract ($n{=}338$).} &
{Excluded ($n{=}271$): off-topic, or software-only work without a
dedicated hardware component. Records remaining: $n{=}67$.} \\
{\textbf{Eligibility}} &
{Full-text articles assessed against the eligibility criteria
($n{=}67$).} &
{Excluded ($n{=}22$): no quantitative hardware result ($n{=}14$);
duplicate publication ($n{=}5$); non-English ($n{=}3$). Records
remaining: $n{=}45$.} \\
\rowcolor{rowgray}
{\textbf{Included}} &
{Studies meeting all inclusion criteria ($n{=}45$).} &
{Backward snowballing added $n{=}18$, mainly analog and
mixed-signal work from 1986--1999, found through the reference lists of
the retained surveys and primary studies.} \\
{\textbf{Final corpus}} &
{\textbf{Studies analyzed in Sections~\ref{sec:taxonomy}
and~\ref{sec:discussion} ($n{=}63$).}} &
{--} \\
\bottomrule
\end{tabular}
\vspace{-0.5em}
\end{table}

\subsection{{Information Sources and Search Strategy}}
\label{subsec:search}

{
Three bibliographic databases were searched: IEEE Xplore, the ACM
Digital Library, and ScienceDirect. The search covers January 2000 to
September 2026. The lower bound reflects the point at which FPGA-based
fuzzy controllers became commercially viable and when the first
microcontroller-class fuzzy engines appeared in the peer-reviewed
literature. Analog and mixed-signal work from the 1980s and early 1990s
predates this window and is poorly indexed by query-based search; such
work was added by backward snowballing from the reference lists of the
retained surveys and is reported separately in Table~\ref{tab:prisma}.
}

{
The query combined three groups of terms with the Boolean operator AND,
and the terms within each group with OR. The \emph{platform group}
comprised \emph{FPGA}, \emph{ASIC}, \emph{VLSI}, \emph{microcontroller},
\emph{MCU}, \emph{embedded}, and \emph{TinyML}. The \emph{technique
group} comprised \emph{fuzzy logic}, \emph{fuzzy inference},
\emph{fuzzy controller}, \emph{membership function}, \emph{stochastic
computing}, and \emph{unary computing}. The \emph{design group}
comprised \emph{hardware implementation}, \emph{hardware accelerator},
\emph{low power}, and \emph{real time}. The field codes and the exact
query strings for each database follow the native syntax of each
platform and are summarized above. The search was restricted to
journal and conference papers, because preprints, theses, and
technical reports are not indexed consistently across the three
databases and their inclusion would have biased the corpus toward one
community.
}

\subsection{{Eligibility Criteria}}
\label{subsec:eligibility}

{
A work was \emph{included} if all three of the following conditions
held:
}
\begin{enumerate}
\item {It reported an original hardware implementation,
      architecture, or hardware-oriented evaluation of a fuzzy
      inference system.}
\item {It provided at least one quantitative result, measured on a
      physical device or reported by a synthesis tool, such as resource
      utilization, latency, throughput, power consumption, energy per
      inference, or inference accuracy.}
\item {It appeared in a peer-reviewed venue.}
\end{enumerate}
{
A work was \emph{excluded} if any of the following conditions held:
}
\begin{enumerate}
\item {It was purely theoretical, with no hardware architecture
      and no hardware-oriented evaluation.}
\item {It addressed software-only implementations of fuzzy
      systems, including those running on general-purpose CPUs or
      GPUs, with no dedicated hardware component.}
\item {It reported no quantitative hardware result, even if it
      described an architecture.}
\item {It duplicated an earlier publication by the same authors
      without additional results, in which case the earlier publication
      was retained.}
\item {It was written in a language other than English.}
\end{enumerate}
{
The full-text exclusion criteria were applied by checking each
candidate against the reported results, not against the abstract; the
abstract often reports accuracy alone and omits the hardware metrics
required by condition~2.
}

\subsection{{Screening and Study Selection}}
\label{subsec:screening}

{
The screening proceeded in three stages, summarized in
Table~\ref{tab:prisma}. After duplicate removal, the remaining records
were screened by title and abstract. Works retained at this stage were
read in full and assessed against the eligibility criteria of
Section~\ref{subsec:eligibility}. The screening was performed by one
reviewer; a random sample of ten percent of the records was
independently re-screened by a second reviewer, and disagreements were
resolved by discussion. The inter-reviewer agreement on the sampled
subset was substantial (Cohen's $\kappa = 0.82$); the sample was not
large enough to estimate agreement with a narrow confidence interval,
and the single-reviewer design is acknowledged as a limitation in
Section~\ref{subsec:method-limitations}.
}

{
Backward snowballing was performed on the reference lists of the
retained surveys and of the retained primary studies. Works added by
snowballing were subjected to the same eligibility criteria as
query-returned records. The snowballing pass was included because a
query-based search does not reliably index conference proceedings from
the 1986--1999 period, during which most of the foundational analog and
mixed-signal fuzzy circuits were published; these works are discussed
in Section~\ref{subsubsec:asic-analog}.
}

\subsection{{Data Coding and Extraction}}
\label{subsec:coding}

{
The retained works were coded along five dimensions:
}
\begin{itemize}
\item {\textbf{Platform}: FPGA device family, ASIC technology
      node, microcontroller family, or analog/mixed-signal process.}
\item {\textbf{Application domain}: control, classification,
      signal processing, sensor interface, or other.}
\item {\textbf{Resource cost}: rule count, membership-function
      count, logic utilization, memory footprint, and area.}
\item {\textbf{Design flow}: manual RTL, high-level synthesis,
      model-based design, or analog custom design.}
\item {\textbf{Evaluation metrics}: latency, throughput, power,
      energy per inference, accuracy, and any reported closed-loop
      quality metric (settling time, overshoot, total harmonic
      distortion).}
\end{itemize}
{
The coded data underpin the tables in Section~\ref{sec:taxonomy} and
the observations and gaps reported in Section~\ref{sec:discussion}.
When a work reported a metric in a form that the coding scheme could
not represent, the metric was recorded as \emph{not reported} rather
than estimated; the resulting density of \emph{N.R.} entries is itself
a finding and is discussed in Section~\ref{sec:discussion}.
}

\subsection{{Reporting Scope}}
\label{subsec:reporting-scope}

{
The tables in Section~\ref{sec:taxonomy} report a representative subset
of the corpus. A work is omitted from the tables when it reports a
single metric that the table columns cannot accommodate, or when its
results are subsumed by a later work from the same group. Works omitted
from the tables are discussed in the text, and the full corpus of
included works is listed in the supplementary material. The subset in
the tables is not a sample selected on outcome; it is selected on the
representability of the metrics it reports, and it is intended to
support the cross-platform comparison of Section~\ref{sec:discussion},
not to substitute for the full corpus.
}

\subsection{{Limitations of the Protocol}}
\label{subsec:method-limitations}

{
Four limitations are noted. First, restricting the search to three
databases may have omitted work published in venues not indexed by
them, in particular regional journals and workshop proceedings; the
snowballing pass partially mitigates this limitation, but it cannot
correct for a systematic coverage bias in the databases themselves.
Second, the exclusion of works without a quantitative hardware result
may have removed contributions that are relevant in principle but
report their evaluation in a form the coding scheme could not capture,
such as a qualitative discussion of a deployed system. This limitation
is not mitigated, and it is reflected in the observation in
Section~\ref{sec:discussion} that the field lacks a reporting standard.
Third, the single-reviewer screening design, described in
Section~\ref{subsec:screening}, introduces the possibility of reviewer
bias in the borderline cases; the sampled re-screening reduces but
does not eliminate this risk. Fourth, the restriction to
English-language publications and to the 2000--2026 window, while
justified above, excludes earlier analog work that was added only by
snowballing, so the pre-2000 corpus is likely less complete than the
post-2000 corpus. These limitations are discussed where they affect
specific conclusions in Sections~\ref{sec:taxonomy}
and~\ref{sec:discussion}.
}

\section{Fuzzy Inference: A Hardware-Oriented Overview}
\label{sec:background}

\begin{figure}[]
	\centering
    \includegraphics[width=\columnwidth]{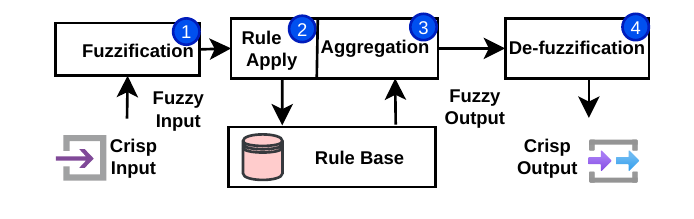}
	\vspace{-1.75em}
	\caption{4-stage inference pipeline of a fuzzy inference system: fuzzification, rule evaluation (rule-apply), aggregation, and defuzzification.}
	\vspace{-1em}
	\label{fig:fig1}
\end{figure}

The translation of fuzzy inference from software to dedicated hardware has been an active research area for over three decades. This effort is driven by a fundamental mismatch: general-purpose processors execute sequentially, while fuzzy rule evaluation is inherently parallel \cite{baturone2000microelectronic, bosque2014fuzzy}. Hardware implementations have evolved across a broad spectrum, from early analog and mixed-signal VLSI designs~\cite{togai1986fuzzy, yamakawa1993fuzzy} and digital fuzzy processors~\cite{dettloff1989fuzzy, eichfeld1996architecture} to modern reconfigurable FPGAs and ultra-low-power MCUs, each offering distinct trade-offs in speed, power, area, and flexibility~\cite{teodorescu2001hardware}.
Bosque \textit{et al.} \cite{bosque2014fuzzy} provide a comprehensive hardware taxonomy spanning two decades, classifying implementations as analog, digital, or mixed, and distinguishing between dedicated ASICs and programmable devices such as FPGAs and {field-programmable analog arrays (FPAAs)}. This evolution has been accompanied by the development of specialized {computer-aided design (CAD)} tools and design methodologies that bridge high-level fuzzy specifications and low-level hardware realization \cite{baturone2000microelectronic}. More recently, the rise of edge computing and TinyML has renewed interest in compact, low-power fuzzy implementations on microcontrollers and SoC platforms, where the trade-off between inference accuracy and resource utilization must be carefully managed \cite{souza2025fuzzy}. The following subsections establish the theoretical foundation from a hardware-centric perspective, bridging fuzzy logic formalism and the physical constraints of digital and analog circuit implementation.

A fuzzy inference system (FIS) maps crisp inputs to crisp outputs through a sequential four-stage process: fuzzification, rule evaluation (rule-apply), aggregation, and defuzzification, as illustrated in Fig.~\ref{fig:fig1} ~\cite{lee1990fuzzy, lee1990fuzzy2, driankov1993fuzzy}.  While several FIS variants, including Mamdani, Sugeno, and Tsukamoto, are available, they share a common architectural template \cite{tang2024fuzzycontrol}.
From a hardware perspective, each of these four stages presents distinct implementation challenges and optimization opportunities. Throughout the pipeline, designers must navigate fundamental trade-offs between precision, latency, area, adaptability, and interpretability \cite{bosque2014fuzzy}.

	\vspace{-1em}
\subsection{Fuzzification Stage}

The first stage of the inference mechanism is \textit{fuzzification}. It converts crisp inputs into membership degrees \(\mu_{A_j}(x_i) \in [0,1]\), representing the degree to which each input belongs to different fuzzy sets. The choice of membership function (MF) shape directly affects hardware cost and accuracy. Early analog implementations often used programmable current-mode circuits for MF generation. In contrast, digital designs have favored {lookup-table (LUT)}-based or piecewise-linear arithmetic approaches \cite{baturone2000microelectronic}.

The most commonly used membership functions (MFs) in hardware are:

\begin{itemize}
\item \textbf{Triangular MF}: Defined as:
\begin{equation}
\mu_{\text{tri}}(x; a, b, c) = 
\begin{cases}
0, & x \leq a \\
\dfrac{x - a}{b - a}, & a < x \leq b \\
\dfrac{c - x}{c - b}, & b < x < c \\
0, & x \geq c
\end{cases}
\label{eq:triangular}
\end{equation}
This MF requires only comparators, adders, and shifters, making it highly efficient in hardware \cite{loukil2020design, abbes2022fuzzy, habeeb2025bell}. 

\item \textbf{Trapezoidal MF}: Extends the triangular form with a flat plateau:
\begin{equation}
\mu_{\text{trap}}(x; a, b, c, d) = 
\begin{cases}
0, & x \leq a \\
\dfrac{x - a}{b - a}, & a < x < b \\
1, & b \leq x \leq c \\
\dfrac{d - x}{d - c}, & c < x < d \\
0, & x \geq d
\end{cases}
\label{eq:trapezoidal}
\end{equation}
This MF offers a flat region of full membership, beneficial for applications requiring insensitivity to small input variations.

\item \textbf{Gaussian MF}: Defined as:
\begin{equation}
\mu_{\text{gauss}}(x; c, \sigma) = \exp\left(-\dfrac{(x - c)^2}{2\sigma^2}\right)
\label{eq:gaussian}
\end{equation}

Gaussian MFs offer smooth transitions but require exponential functions or large LUTs, making them costly in hardware. To address this, researchers have explored alternative MF shapes and implementation strategies that balance accuracy with hardware efficiency. Habeeb and Abbas \cite{habeeb2025bell} demonstrated a 97.7\% slice reduction for Bell-shaped MFs on FPGA through precomputation. In the analog domain, Sefraoui \textit{et al.} proposed a programmable CMOS Bell-shaped MF generator using a differential pair in 45nm technology, achieving 11\,$\mu$W with tunable center and width via $V_{\mathrm{ref}}$ and transistor sizing \cite{sefraoui2025programmable}. Their follow-up work characterized noise in both Bell-shaped and Sigmoid CMOS MF generators, analyzing flicker and thermal noise \cite{sefraoui2026design}. The Sigmoid design achieves a noise floor below 1\,$\mu$V/$\sqrt{\mathrm{Hz}}$ beyond 10\,kHz, with 93\% lower input noise than the Bell-shaped circuit at 10\,kHz, while consuming 12\,$\mu$W. These results confirm its suitability for low-noise analog fuzzy systems.

\item \textbf{Sigmoid MF}: Defined as:
\begin{equation}
\mu_{\text{sigmoid}}(x; a, c) = \dfrac{1}{1 + \exp(-a(x - c))}
\label{eq:sigmoid}
\end{equation}
Sigmoid MFs are particularly useful for classification and pattern recognition applications.
\end{itemize}

From a hardware perspective, the fuzzification stage presents three key design considerations. First, MF shape selection directly impacts arithmetic complexity and resource utilization \cite{bosque2014fuzzy}. Second, MFs can be stored as LUTs in {block RAM (BRAM)} for speed and deterministic latency, or computed using arithmetic circuits for area efficiency \cite{loukil2020design, abbes2022fuzzy, nada2024development}. Third, fuzzification can be fully parallelized for all \(n\) inputs simultaneously, with area scaling as \(O(n \cdot m)\), where \(m\) is the number of MFs per input.

A critical trade-off at this stage is \textit{precision versus resource utilization}. For fixed-point arithmetic with \(I\) integer and \(F\) fractional bits, the quantization step is \(\Delta = 2^{-F}\). Higher precision reduces errors but increases arithmetic unit size by \(O(I+F)\), memory bandwidth, and routing complexity. Studies indicate that 8-bit to 12-bit fixed-point offers an optimal balance for most fuzzy control applications \cite{prabakaran2022fpga}. Recent evaluations of fuzzy control systems on low-cost microcontrollers have shown that defuzzification parameters, such as the iteration step during centroid computation, significantly affect the trade-off between accuracy and processing time. Smaller steps improve precision but can increase computation time beyond 700 ms, while intermediate values offer a favorable balance \cite{souza2025fuzzy}.

\subsection{Rule-Apply (Rule Evaluation) Stage}

Following fuzzification, the \textit{rule-apply} stage (Fig.~\ref{fig:fig1}) evaluates the rule base by applying the fuzzy rules to the membership degrees. In this stage, each rule's antecedent is evaluated to determine its firing strength, and the consequent is then applied to produce an implied fuzzy set. This is where the inference engine performs the core fuzzy reasoning \cite{tang2024fuzzycontrol}.

A generic fuzzy rule \(r\) is expressed as:
\vspace{-0.5em}
\begin{equation}
\begin{split}
\text{IF } & x_1 \text{ is } A_1^r \text{ AND } x_2 \text{ is } A_2^r \text{ AND } \ldots \\
& \text{AND } x_n \text{ is } A_n^r \text{ THEN } y \text{ is } B^r
\end{split}
\label{eq:rule}
\end{equation}

The firing strength of rule \(r\) is computed as:

\begin{equation}
w_r = \mathcal{T}\left(\mu_{A_1^r}(x_1), \mu_{A_2^r}(x_2), \ldots, \mu_{A_n^r}(x_n)\right)
\label{eq:firing_strength}
\end{equation}

where \(\mathcal{T}\) is a t-norm operator. The most common t-norms in hardware are:

\begin{itemize}
\item \textbf{Minimum (Zadeh t-norm)}: \(\mathcal{T}_{\text{min}}(a, b) = \min(a, b)\)  maps to simple comparators and multiplexers, requiring only a few logic gates.
\item \textbf{Algebraic product}: \(\mathcal{T}_{\text{prod}}(a, b) = a \cdot b\)  requires multipliers, consuming significantly more silicon area.
\end{itemize}

From a hardware perspective, the rule-apply stage presents a fundamental latency versus area trade-off: rules can be evaluated in parallel (minimizing latency with area scaling O(R)) or sequentially (minimizing area with latency scaling O(R)). For a system with R rules and n inputs:
\begin{itemize}
\item \textbf{Fully parallel}: L = O(1), A = O(R · n), each rule has dedicated hardware, minimum latency but maximum area.
\item \textbf{Semi-parallel}: L = O(R/P), A = O(P · n), where P is the number of parallel rule evaluation units.
\item \textbf{Fully sequential}: L = O(R), A = O(n), one rule at a time, maximum latency but minimum area.
\end{itemize}
{These asymptotic expressions describe the logical structure of the computation rather than the realized hardware. In practice, a fully parallel design with a large rule base does not attain \(L = O(1)\) after synthesis. Routing delay grows with fan-out and interconnect length, the depth of the aggregation tree adds \(\lceil \log_2 R \rceil\) levels, and memory access to the rule base becomes a bottleneck once the design exceeds the available block-RAM ports. A synthesis-aware analysis therefore replaces the idealized \(L = O(1)\) of the parallel form with a small constant that grows with \(R\), and the idealized \(A = O(R \cdot n)\) with an area that grows superlinearly once routing and buffering are counted. The three forms above remain useful as a first-order guide, but the latency and area of a synthesized design must be measured rather than inferred from these expressions.}

Early ASIC implementations, such as the fuzzy processor by Dettloff \textit{et al.} (1989), achieved 580 KFLIPS with 51 rules in 1$\mu$m CMOS, while more recent designs have demonstrated significant improvements in both speed and integration density \cite{bosque2014fuzzy}. {At the circuit level,} the min operation maps efficiently to comparators while product requires multipliers; and firing strengths \(w_r\) for all \(R\) rules must be stored in registers or memory buffers for subsequent aggregation \cite{talaska2019parallel, talaska2023novel}. The rule-apply operation can be highly parallelized in hardware with each rule assigned to dedicated logic, enabling all rules to fire simultaneously{, a key advantage over software implementations}.

\subsection{Aggregation and Implication}

In the implication stage, which follows the rule-apply stage, each rule's output is modified based on its firing strength:

\begin{equation}
\mu_{B^r}'(y) = \mathcal{T}\left(w_r, \mu_{B^r}(y)\right)
\label{eq:implication}
\end{equation}

The aggregation stage (Fig.~\ref{fig:fig1}) then combines all rule outputs into a single fuzzy set:

\begin{equation}
\mu_B'(y) = \mathcal{S}\left(\mu_{B^1}'(y), \mu_{B^2}'(y), \ldots, \mu_{B^R}'(y)\right)
\label{eq:aggregation}
\end{equation}

where \(\mathcal{S}\) is typically the maximum operator: \(\mathcal{S}_{\text{max}}(a, b) = \max(a, b)\). From a hardware perspective, aggregation is highly parallelizable using a tree of max operators with depth \(\lceil \log_2 R \rceil\), significantly reducing latency compared to sequential accumulation \cite{swami2022fpga}. The choice between min and product implication has significant hardware implications: product requires multipliers, while min requires only comparators \cite{koko2025new}.

This stage also highlights the \textit{adaptability versus fixed-function efficiency} trade-off. Fixed-function systems with predetermined MFs and rule bases achieve highest efficiency through specialized datapaths and ROM-based storage \cite{loukil2020design, abbes2022fuzzy}. Adaptive systems{, such as neuro-fuzzy and the Adaptive Neuro-Fuzzy Inference System (ANFIS), require} writable memory (SRAM), multipliers for gradient computation, and additional control logic, substantially increasing hardware cost \cite{lin2025implementation, hermassi2024zynq, jiang2026memristive}. 

\subsection{Defuzzification Stage}

The final stage shown in Fig.~\ref{fig:fig1} is the \textit{defuzzification process}, which converts the aggregated fuzzy set to a crisp (non-fuzzy) output value{, widely recognized as the} \textit{computational bottleneck} in fuzzy hardware \cite{odry2021stochastic}. The most common defuzzification methods are:

\begin{itemize}
\item \textbf{Centroid (Center of Gravity)}: 
\begin{equation}
y^*_{\text{centroid}} = \left( \textstyle\sum_{k=1}^{K} y_k \cdot \mu_B'(y_k) \right) \big/ \left( \textstyle\sum_{k=1}^{K} \mu_B'(y_k) \right)
\label{eq:centroid_discrete}
\end{equation}
This requires multiplication, accumulation, and division, all costly operations in hardware that consume significant silicon area and power. Souza \textit{et al.} \cite{souza2025fuzzy} have shown that the iteration step during centroid computation critically affects the trade-off between precision and processing time on low-cost microcontrollers.

\item \textbf{Weighted Average (Sugeno)}:
\begin{equation}
y^*_{\text{sugeno}} = \left( \textstyle\sum_{r=1}^{R} w_r \cdot f^r(\mathbf{x}) \right) \big/ \left( \textstyle\sum_{r=1}^{R} w_r \right)
\label{eq:sugeno}
\end{equation}
Similarly requiring division hardware \cite{pavitra2025design}.

\item \textbf{Approximate methods}: To reduce hardware complexity, designers employ maximum membership principle (no division, lowest accuracy), LUT pre-computation of normalized outputs (fast but memory-intensive), iterative division algorithms (\textit{e.g.,} non-restoring or Newton-Raphson), or stochastic and unary bit-stream-based methods (Section~\ref{subsubsec:fpga-stochastic}) at the cost of some accuracy \cite{jalilvand2020fuzzy, 9969358}.
\end{itemize}

The defuzzification stage also exemplifies the \textit{interpretability versus complexity} trade-off. While the inherent interpretability of fuzzy logic, where each rule corresponds to an explicit linguistic statement, is a key advantage, the hardware complexity grows with rule base size. For \(n\) inputs and \(m\) MFs per input, \(R_{\text{max}} = m^n\), leading to exponential rule growth. Type-2 fuzzy systems further increase complexity by modeling uncertainty intervals \(\mu_A(x) \in [\underline{\mu}_A(x), \overline{\mu}_A(x)]\), requiring iterative type-reduction algorithms that challenge real-time hardware implementation \cite{matusi2022fpga}.

\section{Platform-Based Taxonomy of Fuzzy Hardware Implementations}
\label{sec:taxonomy}

This section provides a detailed examination of the three principal platform categories: FPGA-based, ASIC-based, and embedded/TinyML implementations. The four pipeline stages of Section~\ref{sec:background} map differently onto each substrate: FPGAs expose parallel rule paths through LUTs, digital signal processing (DSP) slices, and BRAM; ASICs trade reconfigurability for custom datapaths and tighter performance--power--area (PPA); embedded targets fold the same stages into fixed-point software or small accelerators under kilobyte-scale memory budgets. The following subsections examine how each platform organizes these stages, maps them onto its resources, and manages the resulting design trade-offs.

\subsection{FPGA-Based Hardware Implementations}
\label{subsec:fpga}

FPGAs are widely used for fuzzy hardware acceleration. They balance performance, flexibility, and time-to-market. Unlike fixed-function ASICs, FPGAs can be reconfigured to implement different fuzzy architectures, rule bases, and precision levels without incurring the high non-recurring engineering costs of custom silicon fabrication. This reconfigurability suits research prototyping, adaptive systems, and applications with evolving requirements \cite{loukil2020design, abbes2022fuzzy}.

Modern FPGA devices integrate a rich set of resources that directly support fuzzy inference. These include LUTs for combinatorial logic, flip-flops for state storage, DSP slices for high-performance arithmetic (multipliers, adders, and accumulators), BRAMs for memory-intensive operations such as membership function storage, and high-speed transceivers for I/O-intensive applications \cite{swami2022fpga, koko2025new}. The systematic mapping of fuzzy computations to these resources, and the trade-offs among them, constitutes the central challenge in FPGA-based fuzzy system design.

A typical FPGA-based fuzzy implementation comprises the following hardware modules:
\begin{itemize}
\item \textbf{Input interface and synchronization logic}: Sampling and synchronizing multiple input channels.
\item \textbf{Membership function generators (MFGs)}: Often implemented as LUTs in BRAM or as piecewise-linear arithmetic circuits \cite{nada2024development}.
\item \textbf{Rule evaluation array}: A set of parallel or semi-parallel rule blocks, each containing antecedent evaluation and consequent composition logic \cite{de2024fpga}.
\item \textbf{Aggregation and defuzzification unit}: Accumulating rule outputs and computing the final crisp value.
\item \textbf{Output interface}: Driving actuators, displays, or communication links.
\end{itemize}

Table~\ref{tab:fpga_summary} provides a {summary} of representative FPGA-based fuzzy implementations across all architecture categories. The following subsections discuss each category in detail.

\begin{table*}[]
\centering
\caption{Summary of FPGA-Based Fuzzy Logic Implementations. Abbreviations: {FLC} = fuzzy logic controller; {THD} = total harmonic distortion; {DTC} = direct torque control; {GWO} = grey wolf optimization; {XSG} = Xilinx System Generator; {CNN} = convolutional neural network; {FLDTC} = fuzzy logic-based direct torque control; {CRFNN} = convolutional recurrent fuzzy neural network; {IT2FLS} = interval type-2 fuzzy inference system; {FNN} = fuzzy neural network.}
\label{tab:fpga_summary}
\renewcommand{\arraystretch}{1.3}
\setlength{\tabcolsep}{5pt}
\footnotesize
\begin{tabular}{@{} 
    >{\centering\arraybackslash}m{0.5cm}   
    >{\raggedright\arraybackslash}m{2.8cm} 
    >{\raggedright\arraybackslash}m{3.4cm} 
    >{\raggedright\arraybackslash}m{2.4cm} 
    >{\centering\arraybackslash}m{1.0cm}   
    >{\raggedright\arraybackslash}m{6.0cm} 
    @{}}
\toprule
\rowcolor{headerblue}
{\color{white}\textbf{Ref.}} &
{\color{white}\textbf{Application}} &
{\color{white}\textbf{Architecture}} &
{\color{white}\textbf{FPGA Platform}} &
{\color{white}\textbf{Rules}} &
{\color{white}\textbf{Key Contribution}} \\
\midrule

\cite{loukil2020design} & MPPT (PV) & INC-Fuz (Sugeno) & Stratix III & 25 & 4436 PV channels, 0.1127 ms exec. time \\
\rowcolor{rowgray}
\cite{abbes2022fuzzy} & MPPT (Multi-ch. PV) & Fuzzy (variable step) & Stratix III & 25 & Distributed MPPT, partial shading tolerance \\
\cite{swami2022fpga} & DTC (IM drive) & FLDTC & Spartan XC3S1400AN & 49 & Reduced torque ripple, 0.5 rad/s undershoot \\
\rowcolor{rowgray}
\cite{meghwal2024robust} & Matrix converter IM drive & Robust FLC & FPGA & 5 & Robust control, 2.4\% THD, FPGA validation \\
\cite{koko2025new} & Shunt active filter & FLC (trapezoidal) & Altera Cyclone V & 9 & THD 22.65\% $\to$ 4.79\%, IEEE 519 \\
\rowcolor{rowgray}
\cite{nada2024development} & Mechatronic control & Sugeno (zero-order) & sbRIO-9631 & 25 & 2\% overshoot, 0.85 s settling time \\
\cite{ahmed2024implementation} & DC motor control & FLC + GWO & {XSG} & 9 & MF optimization, 8,132 LUTs \\
\rowcolor{rowgray}
\cite{prabakaran2022fpga} & Productivity prediction & FLC & Zynq xc7z010 & 15 & 96\% accuracy, 472.13 MHz \\
\cite{hermassi2024zynq} & Wind turbine MPPT & TS-ANFIS & Zynq Zed-Board & 27 + 4 & 99.84\% efficiency, 0.215 W \\
\rowcolor{rowgray}
\cite{lin2025implementation} & Human activity recognition & CRFNN & ZCU104 & 4 & 95.05\% accuracy, 4.78$\times$ faster than GPU \\
\cite{jiang2026memristive} & Classification & Memristive FNN & Programmable circuit & 16 & Full-analog CIM, 1.35$\times$10$^5$ faster than CPU \\
\rowcolor{rowgray}
\cite{matusi2022fpga} & Quadrotor attitude control & IT2FLS + Nie-Tan & EP4CE115F29C7 & 16 & 2.38--4.77 Mega FLIPS, 4\% logic \\
\cite{jalilvand2020fuzzy} & Fuzzy inference engine & Unary computing & Virtex-7 & 81 & 94\% LUT reduction, 82\% area saving \\
\rowcolor{rowgray}
\cite{odry2021stochastic} & Self-balancing robot control & Stochastic computing & FPGA & 9 & First experimental validation, noise-tolerant \\
\cite{9969358} & Image noise reduction & Stochastic computing & Virtex-7 & 16 & 91.8\% area saving, 84.7\% power reduction \\
\rowcolor{rowgray}
\cite{srivastava2026reconfigurable} & Drone detection & YOLO + CNN + Fuzzy Fusion & Kria KV260 + Zedboard & 9 & 97\% accuracy, 3 ms response, 10\% power saving \\
{\cite{farhadi2026scfnns}} & {Classification (Iris, Wine, Breast)} & {Stochastic FNN} & {XA Artix UltraScale+} & {N.R.} & {36--53\% LUT reduction, DSP-free, 97\% faster training} \\
\bottomrule
\end{tabular}%
\vspace{-1em}
\end{table*}

\subsubsection{General-Purpose FPGA Fuzzy Controllers}
\label{subsubsec:fpga-general}

Conventional FPGA implementations of fuzzy controllers emphasize high-speed, deterministic inference for real-time control applications. These designs typically employ \textit{parallel rule evaluation}, where each rule is assigned to dedicated hardware resources, enabling all rules to be evaluated concurrently within a single clock cycle or a small number of pipelined stages \cite{de2024fpga}. The membership functions are either implemented as BRAM-based LUTs for maximum speed or as arithmetic units for area efficiency \cite{nada2024development}.

\begin{figure}[]
\centering
\includegraphics[width=0.95\columnwidth,trim=0.6cm 0.5cm 0.5cm 0cm,clip]{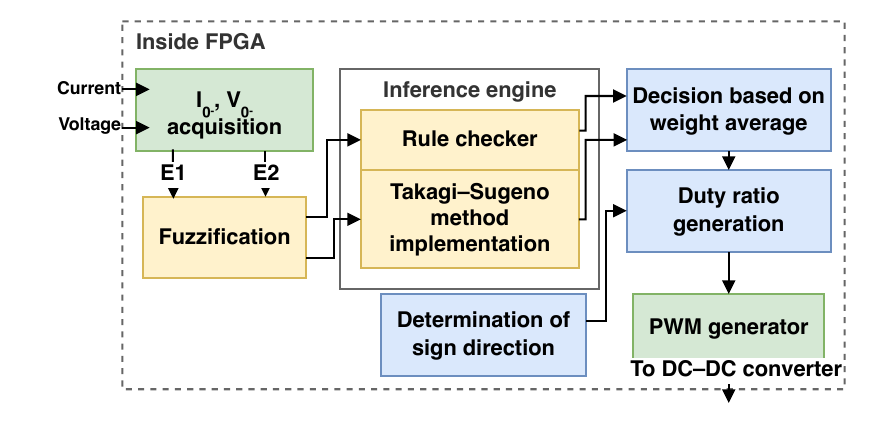}
\vspace{-1.em}
\caption{Hardware organization of the Sugeno-type fuzzy MPPT controller inside the FPGA. The $I_{pv}$/$V_{pv}$ acquisition front-end (green) generates the error signals E1/E2 for the fuzzification unit, whose membership degrees drive a rule checker and a Takagi--Sugeno implementation block within the inference engine (yellow); a weight-average decision unit and the duty-ratio generation logic (blue), assisted by the sign-direction detector, feed the PWM generator (green) that commands the external DC--DC converter. The dashed boundary marks the resources mapped onto the FPGA fabric \cite{loukil2020design}.}
\label{fig:loukil_fpga}
\vspace{-1.5em}
\end{figure}

Several representative works illustrate the capabilities of this approach. In the domain of renewable energy, FPGA-based fuzzy controllers for {MPPT} in photovoltaic systems have been extensively studied \cite{loukil2020design, abbes2022fuzzy}. Loukil \textit{et al.} \cite{loukil2020design} propose a hybrid INC-Fuz algorithm that combines incremental conductance with fuzzy logic, implementing a Sugeno-type fuzzy system with 25 rules on a Stratix III FPGA. \textcolor{black}{The hardware organization of this controller within the FPGA is illustrated in Fig.~\ref{fig:loukil_fpga}, which details the acquisition front-end, the fuzzification and inference stages, and the {pulse-width modulation (PWM)} generation logic.} The design achieves an execution time of 0.1127 ms and demonstrates the ability to control up to 4436 PV channels, outperforming microcontroller and DSP-based implementations by a factor of 300. Building upon this work, Abbes \textit{et al.} \cite{abbes2022fuzzy} extend the approach to multi-channel photovoltaic systems, employing a distributed MPPT architecture where each PV panel operates at its own maximum power point. Their FPGA-based implementation achieves significant reductions in computation time while maintaining stable operation under partial shading conditions.

Similarly, motor control applications have benefited from FPGA acceleration. Swami \textit{et al.} \cite{swami2022fpga} present an {FLDTC} scheme for induction motor drives fed by matrix converters. The design employs a fuzzy logic controller with 49 rules (seven membership functions for each of the two inputs) to adaptively adjust the torque hysteresis band. Implemented on a Xilinx Spartan XC3S1400AN FPGA, the FLDTC scheme significantly reduces electromagnetic torque ripple and improves both steady-state and dynamic response compared to conventional DTC. Experimental results demonstrate a speed undershoot of only 0.5 rad/s during step changes, compared to 3.2 rad/s for conventional DTC. Meghwal \textit{et al.} \cite{meghwal2024robust} further explore robust fuzzy controller design for matrix converter-fed induction motor drives, addressing parameter uncertainties and load variations through FPGA-based implementation with a Mamdani-type FLC. Their design achieves a stator current THD of 2.4\% and demonstrates stable operation under variable voltage and frequency conditions.

\begin{figure}[]
\centering
\includegraphics[width=0.95\columnwidth,trim=0.4cm 0cm 0.4cm 0.2cm,clip]{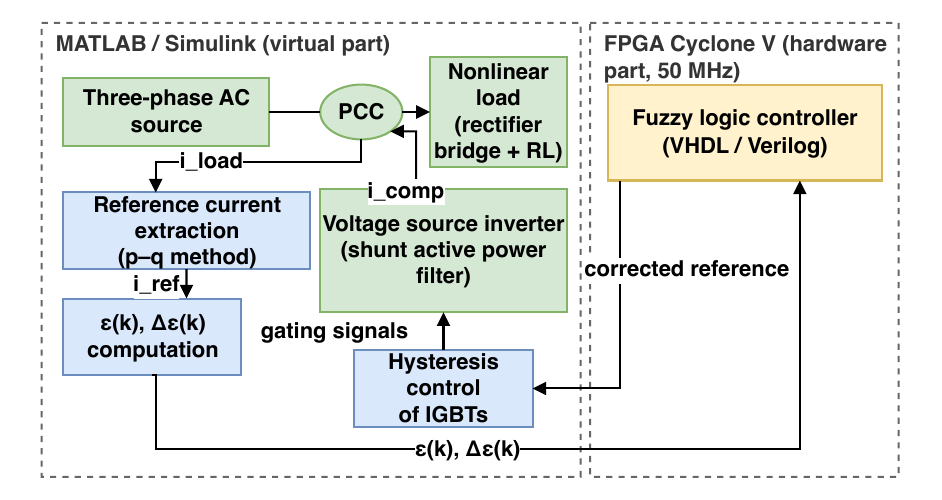}
\vspace{-1.em}
\caption{Block diagram of the hardware-in-the-loop co-simulation platform for the fuzzy-controlled shunt active power filter. The virtual part in MATLAB/Simulink (dashed region, left) models the three-phase AC source, the nonlinear rectifier--RL load, the p--q theory reference-current extraction, and the hysteresis switching of the voltage-source inverter that injects the compensation current $i_{comp}$ at the point of common coupling (PCC); the hardware part (dashed region, right) hosts the fuzzy logic controller (yellow) described in VHDL/Verilog on an Altera Cyclone~V device running at 50~MHz. The FLC receives the current error $\varepsilon(k)$ and its variation $\Delta\varepsilon(k)$ and returns the corrected compensation reference through the JTAG-based HIL link (dashed connection); green blocks denote the power-system elements and blue blocks the signal-processing stages\cite{koko2025new}.}
\label{fig:koko_cosim}
\end{figure}

The use of fuzzy logic in power electronics has also been extended to power quality improvement. Koko \textit{et al.} \cite{koko2025new} present an FPGA-based fuzzy logic controller for shunt active power filters, implemented on an Altera Cyclone V FPGA with a trapezoidal membership function design and 9 inference rules. \textcolor{black}{The {hardware-in-the-loop (HIL)} co-simulation platform for this system is shown in Fig.~\ref{fig:koko_cosim}, illustrating the virtual MATLAB/Simulink environment, the FPGA-based fuzzy controller, and the JTAG communication link between them.} Through HIL co-simulation, the design achieves a reduction in THD from 22.65\% to 4.79\%, meeting IEEE 519 standards. The implementation utilizes 26\% of available LUTs and consumes all available DSP blocks, demonstrating the resource efficiency of FPGA-based fuzzy control for power electronics applications.

Beyond power electronics and renewable energy, FPGA-based fuzzy controllers have been deployed in robotics, automotive systems, and industrial automation. Nada and Bayoumi \cite{nada2024development} present a comprehensive framework for implementing PD-like and {proportional--integral--derivative (PID)}-like fuzzy controllers on NI sbRIO-9631 platforms using LabVIEW FPGA. Their design employs a zero-order Sugeno fuzzy system with five triangular membership functions for each of two inputs and 25 rules. Through optimization using genetic algorithms, the optimized PID-FLC-FPGA controller achieves an overshoot of 2\% and settling time of 0.85 s, significantly outperforming conventional PID-FPGA controllers which exhibit 32\% overshoot and 1.5 s settling time.

Optimization techniques have also been applied to improve the performance of FPGA-based fuzzy controllers. Ahmed and Yahia \cite{ahmed2024implementation} employ {GWO} to optimize membership functions and rule bases for a geared DC motor speed controller. Implemented using {XSG} and co-simulated on FPGA, the optimized controller reduces computational complexity while maintaining control accuracy. The design achieves resource utilization of 8,132 LUTs (15\%) and 149 DSP slices (67\%) on a Xilinx FPGA target.

Prabakaran \textit{et al.} \cite{prabakaran2022fpga} demonstrate FPGA-based fuzzy logic for productivity prediction in intelligent embedded systems, showcasing the versatility of fuzzy controllers beyond conventional control applications. Their intelligent embedded fuzzy decision support system (IEFDSS), implemented on a Zynq xc7z010 FPGA, achieves 96\% accuracy with a 472.13 MHz operating frequency, demonstrating the potential of FPGA-based fuzzy systems for agricultural decision support.

The common theme across these implementations is the emphasis on deterministic timing, scalable parallelism, and fixed-point optimization. A key challenge remains managing resource growth as the rule base expands, with techniques such as rule clustering, hierarchical inference, and partial reconfiguration proposed to address this limitation \cite{bartok2022design}.

\subsubsection{FPGA-Based Neuro-Fuzzy and ANFIS Architectures}
\label{subsubsec:fpga-neurofuzzy}

Neuro-fuzzy systems {couple} the learning capabilities of neural networks with the interpretability of fuzzy logic, with {ANFIS} being the most prominent and widely studied architecture. Implementing ANFIS and related neuro-fuzzy models on FPGA presents unique challenges beyond those of standard fuzzy controllers, as the hardware must support not only inference but also \textit{learning}{, namely the iterative adjustment} of membership function parameters and rule consequents based on training data.
The primary design considerations for FPGA-based neuro-fuzzy systems include dual-mode operation (inference and learning), nonlinear function support (sigmoids, Gaussians), and high memory bandwidth for parameter storage and gradient buffers.

\begin{figure}[]
\centering
\includegraphics[
    width=\columnwidth,
    page=1,
    trim=1.1cm 0.3cm 1.1cm 0.4cm,
    clip
]{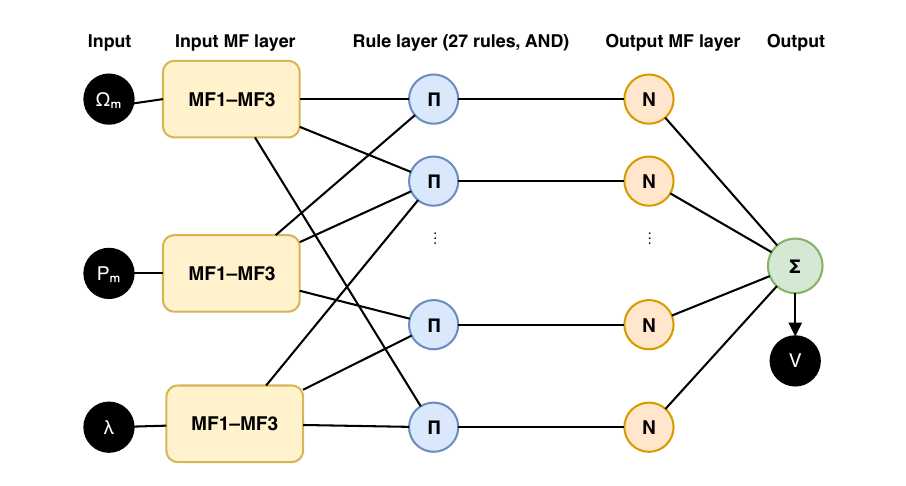}
\vspace{-1em}

\caption{TS-ANFIS. Top: layered structure of the WSE TS-ANFIS model, where the three inputs $\Omega_m$, $P_m$, and $\lambda$ (TSR) pass through three triangular membership functions each (yellow), a 27-rule product layer ($\Pi$), and a normalization layer (N, orange) before summation ($\Sigma$, green) yields the estimated wind speed $V$ \cite{hermassi2024zynq}.}
\label{fig:hermassi_anfis}
\vspace{-1.5em}
\end{figure}

Hermassi \textit{et al.} \cite{hermassi2024zynq} present a {detailed} FPGA implementation of a Takagi-Sugeno ANFIS for {MPPT} in a grid-connected wind turbine system. \textcolor{black}{The layered architecture of their wind speed estimator (WSE) TS-ANFIS model is shown in Fig.~\ref{fig:hermassi_anfis}, which illustrates how the three inputs{ (mechanical speed, mechanical power, and tip speed ratio)} pass through triangular membership functions, a 27-rule product layer, and normalization before summation yields the estimated wind speed.} Their work introduces two key ANFIS-based components: {the WSE}, which eliminates the need for physical anemometers, and an MPPT controller that maximizes power extraction. The WSE-based TS-ANFIS model employs three triangular membership functions for each of its three inputs: mechanical speed, mechanical power, and tip speed ratio (TSR){, generating} 27 fuzzy rules. Trained using a hybrid optimization method over 200 epochs, the estimator achieves a {root mean square error (RMSE)} of 0.0085381 and R² of 0.99985. The MPPT-based TS-ANFIS controller achieves 99.84\% tracking efficiency, significantly outperforming conventional PID controllers (92.63\%). Implemented on a Xilinx Zynq Zed-Board using XSG, the design achieves 80.38 MHz, 2572.16 Mbps throughput, and 0.215 W power consumption, utilizing 46.6\% of LUTs and 17.5\% of flip-flops.

\begin{figure*}[]
\centering
\includegraphics[width=1.6\columnwidth,page=1,trim=0.5cm 0cm 0.5cm 0cm,clip]{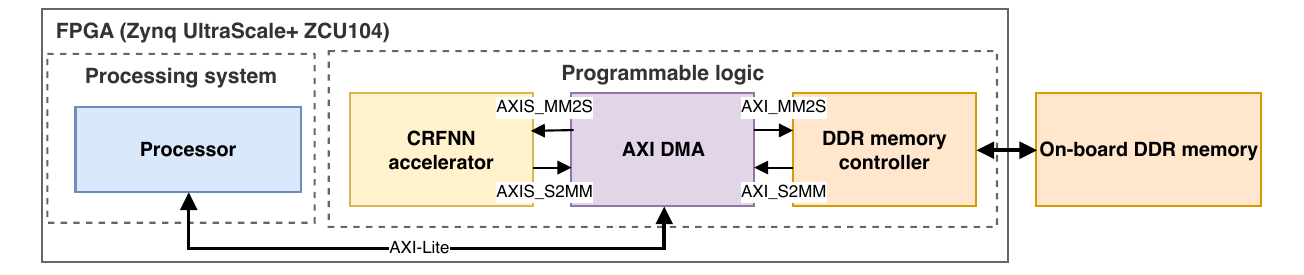}
\vspace{-0.5em}
\caption{Hardware architecture of the CRFNN accelerator for real-time human activity recognition. Top: system-level view on the Zynq UltraScale+ ZCU104, where the processing system (dashed) configures the design over AXI-Lite while the programmable logic hosts the CRFNN accelerator (yellow), the AXI DMA engine (purple) with its AXIS\_MM2S/AXIS\_S2MM streaming channels, and the DDR memory controller (orange) that connects to the on-board DDR memory over a bidirectional bus (thick lines) \cite{lin2025implementation}.}
\label{fig:lin_crfnn}
\vspace{-1em}
\end{figure*}

Lin \textit{et al.} \cite{lin2025implementation} demonstrate a {CRFNN} on FPGA for real-time human activity recognition, achieving state-of-the-art accuracy with sub-millisecond inference latency. \textcolor{black}{The system-level hardware architecture of this CRFNN accelerator on the Zynq UltraScale+ ZCU104 platform is depicted in Fig.~\ref{fig:lin_crfnn}, showing the processing system, the programmable logic hosting the CRFNN accelerator, the AXI DMA engine with streaming channels, and the DDR memory controller.} Implemented on a ZCU104 FPGA, the CRFNN achieves 95.05\% accuracy on the UCI-HAR dataset and 97.91\% on the WISDM dataset. The FPGA implementation achieves 4.78× faster inference speed and 11.17× lower energy consumption compared to a GPU, demonstrating the effectiveness of FPGA acceleration for embedded HAR applications. The design utilizes 7.47\% LUTs, 7.09\% FFs, 22.92\% BRAMs, and 10.01\% DSPs on the ZCU104 platform.

A key challenge in FPGA-based neuro-fuzzy systems is the dichotomy between offline and online learning. While offline learning (training in software, deployment in hardware) is well-established, online learning{, where the hardware adapts its parameters during operation, remains} largely unexplored due to the computational demands of gradient computation and parameter updates \cite{hermassi2024zynq}. This limitation is particularly acute in resource-constrained edge scenarios, where retraining from scratch is impractical.

Jiang \textit{et al.} \cite{jiang2026memristive} explore an innovative direction where memristive crossbars are used to implement synaptic weights and fuzzy membership functions in a programmable circuit system, blurring the boundary between conventional FPGA logic and emerging non-volatile memory technologies for neuro-fuzzy computation. Their work presents the first full-analog circuit scheme for FNN based on computing-in-memory architectures, achieving 1.35×10$^5$× faster inference speed, 914× smaller area, and 33× lower power consumption compared to CPU and FPGA implementations.

Despite these advances, FPGA-based neuro-fuzzy systems remain challenging to design and optimize due to the trade-off between learning capability and resource utilization, and the lack of standardized high-level design flows.

\subsubsection{FPGA Implementations of Type-2 and Interval Type-2 Fuzzy Systems}
\label{subsubsec:fpga-type2}

Type-2 fuzzy logic systems extend traditional type-1 FLS by modeling uncertainty in the membership functions themselves, providing enhanced robustness to noise and parameter variations. However, this comes at the cost of significantly higher computational complexity, particularly in the \textit{type-reduction} stage.

\begin{figure*}[]
\centering
\includegraphics[
    width=1.4\columnwidth,
    page=2,
    trim=0.5cm 0cm 0.5cm 0cm,
    clip
]{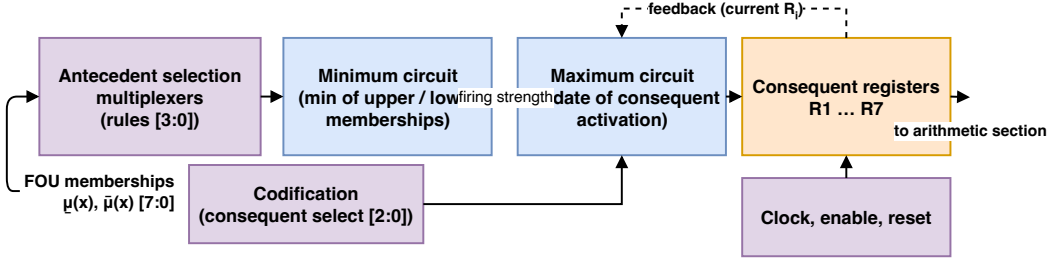}
\vspace{-1em}
\caption{Internal organization of the sequenced inference engine of the IT2FLS processor. Rule-indexed multiplexers (purple) select the appropriate upper/lower FOU membership values $\underline{\mu}(x)$, $\overline{\mu}(x)$; a minimum circuit (blue) computes the rule firing strength, and a maximum circuit (blue) updates the corresponding consequent register R1--R7 (orange) selected by the codification unit (purple), with the dashed path denoting the current register value fed back for the max comparison. Clock, enable, and reset are distributed by the control block, and the accumulated consequent activations are forwarded to the Nie--Tan arithmetic section \cite{matusi2022fpga}.}
\label{fig:matusi_ie}
\vspace{-1em}

\end{figure*}

Matusi \textit{et al.} \cite{matusi2022fpga} present an FPGA implementation of an {IT2FLS} for quadrotor attitude control. \textcolor{black}{The internal organization of the sequenced inference engine of this IT2FLS processor is illustrated in Fig.~\ref{fig:matusi_ie}, which details how rule-indexed multiplexers select upper/lower {footprint of uncertainty (FOU)} membership values, a minimum circuit computes rule firing strengths, and a maximum circuit updates the corresponding consequent registers.} Instead of a fully parallel inference engine, the design utilizes a rule base management unit that sequences active rules through a single inference engine circuit, significantly reducing hardware resource utilization. A key innovation is the use of the Nie-Tan (NT) type-reduction algorithm, which performs both type-reduction and defuzzification in a single step:

\begin{equation}
Y_{NT} = \frac{\sum_{i=1}^{M} y^i (\bar{f}^i + \underline{f}^i)}{\sum_{i=1}^{M} (\bar{f}^i + \underline{f}^i)}
\label{eq:nie_tan}
\end{equation}

The IT2FLS processor, implemented on an Altera EP4CE115F29C7 FPGA with two 8-bit inputs, four Gaussian MFs per input, and sixteen rules, achieves 14.3 MHz operating frequency, consuming only 4\% of logic elements, and delivers 2.38 to 4.77 Mega FLIPS{, which is sufficient for real-time control applications}.

\subsubsection{FPGA-Based Stochastic and Unary Fuzzy Architectures}
\label{subsubsec:fpga-stochastic}

Stochastic computing (SC)~\cite{Najafi_TVLSI_2019} and unary computing (UC) trade precision for lower hardware cost and power consumption. In this unconventional method of computing, numeric values are represented as uniform bit-streams, and arithmetic operations are reduced to simple logic gates.

\begin{figure}[]
\centering
\includegraphics[
    width=\columnwidth,
    page=1,
    trim=0.8cm 0.2cm 0.8cm 0cm,
    clip
]{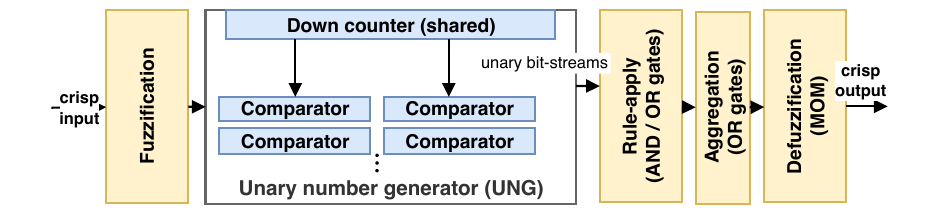}
\caption{General structure of the unary bit-stream fuzzy inference engine. Crisp inputs are fuzzified (yellow) and converted into unary bit-streams by the unary number generator (UNG), in which a single down-counter (blue) is shared among parallel comparators to minimize generation overhead; the rule-apply and aggregation stages (yellow) are then realized with simple \texttt{AND}/OR gates operating directly on the bit-streams, and a Middle-of-Maxima (MOM) defuzzifier produces the crisp output \cite{jalilvand2020fuzzy}.}
\label{fig:jalilvand_unary}
\vspace{-1em}
\end{figure}

\begin{figure*}[]
\centering
\includegraphics[
    width=1.7\columnwidth,
    trim=0.5cm 0.5cm 0.5cm 0cm,
    clip
]{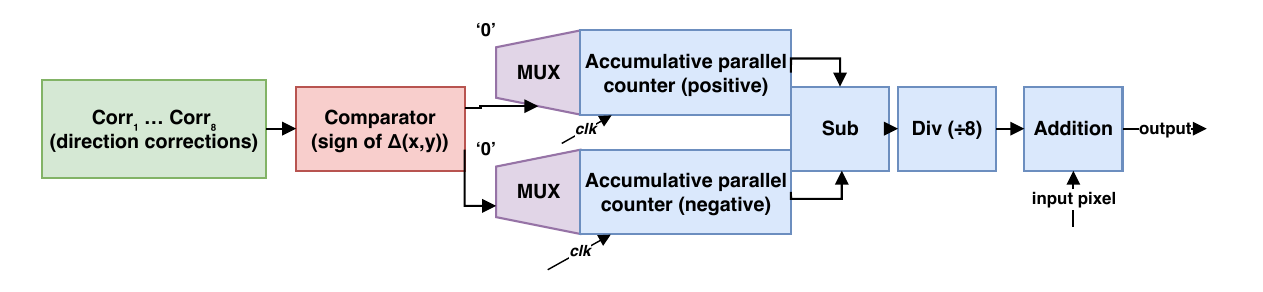}
\vspace{-1.25em}
\caption{Stochastic implementation of the correction accumulator. The eight directional correction streams Corr$_1$\,\ldots\,Corr$_8$ (green) are split by a sign comparator (red) into positive and negative subsets through two multiplexers (purple); two accumulative parallel counters (blue) implicitly convert the bit-streams back to binary, and their difference is divided by eight and added to the original input pixel to yield the denoised output, replacing costly binary multipliers and dividers with counting logic \cite{9969358}.}
\label{fig:estiri_acc}
\vspace{-0.75em}
\end{figure*}

Jalilvand \textit{et al.} \cite{jalilvand2020fuzzy} demonstrate fuzzy logic operations using unary bit-streams on FPGA, \textcolor{black}{achieving 94\% reduction in LUT utilization and 82\% area savings on a Virtex-7 FPGA with up to 81 rules, as illustrated in the general structure shown in Fig.~\ref{fig:jalilvand_unary}, where crisp inputs are fuzzified, converted to unary bit-streams by a shared unary number generator, and processed through \texttt{AND}/\texttt{OR} gates before a Middle-of-Maxima defuzzifier produces the crisp output.} Odry \textit{et al.} \cite{odry2021stochastic} present one of the first experimental validations of a stochastic logic-based fuzzy logic controller on FPGA, demonstrating real-time self-balancing robot control with inherent noise tolerance. Estiri \textit{et al.} \cite{9969358} propose a SC-based fuzzy filter for image noise reduction, \textcolor{black}{achieving 91.8\% area savings and 84.7\% power reduction on Virtex-7 FPGA through the stochastic implementation of the correction accumulator shown in Fig.~\ref{fig:estiri_acc}, where directional correction streams are split by sign comparators and processed by accumulative parallel counters.}
The primary advantages include extremely low hardware cost, inherent fault tolerance, and simplified defuzzification. Limitations include convergence time, bit-stream generation overhead, and accuracy degradation. 

\subsubsection{FPGA-Based Fuzzy Systems on SoC and Reconfigurable Edge Platforms}
\label{subsubsec:fpga-soc}

{Finally, heterogeneous SoCs combine a software processor with reconfigurable logic, so inference can split between firmware and dedicated fuzzy accelerators on the same die.} The integration of embedded processors with programmable logic in modern FPGA devices has given rise to heterogeneous \textit{{SoC}} platforms, enabling hardware-software co-design for edge AI applications.

\begin{figure}[]
\centering
\includegraphics[
    width=0.9\columnwidth,
    trim=0.cm 0.cm 0.cm 0.cm,
    clip
]{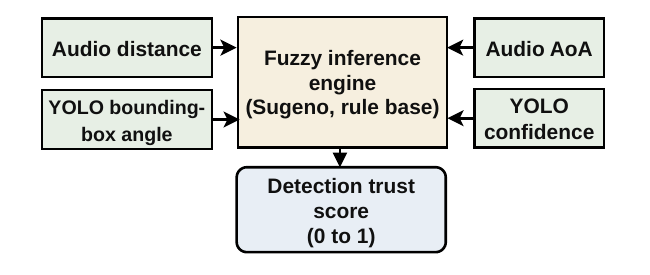}
\vspace{-0.75em}
\caption{Fuzzy fusion framework of the drone-detection system. The four perceptual inputs (green), audio angle of arrival, audio distance, YOLO confidence, and YOLO bounding-box angle, are fuzzified and combined by the 9-rule Sugeno fuzzy inference engine (yellow), which produces a DTS between 0 (unreliable) and 1 (highly reliable) used for the final multi-modal decision (blue). \cite{srivastava2026reconfigurable}}
\label{fig:srivastava_fusion}
\vspace{-1em}
\end{figure}

Srivastava \textit{et al.} \cite{srivastava2026reconfigurable} present a reconfigurable fuzzy-logic audio-visual fusion system for drone detection. {As shown in Fig.~\ref{fig:srivastava_fusion}, a 9-rule Sugeno engine fuses four perceptual inputs (audio angle of arrival, audio distance, YOLO confidence, and YOLO bounding-box angle) into a Detection Trust Score (DTS) in $[0,1]$, combining YOLO11 visual detection with a lightweight CNN for audio classification.} Implemented on AMD Kria KV260 and Zedboard, the system achieves over 97\% detection accuracy, with the fuzzy motor controller achieving 3 ms response time (40\% faster than PID) and 10\% power reduction. {The Zynq-based TS-ANFIS implementation of Hermassi \textit{et al.}~\cite{hermassi2024zynq}, introduced in Section~\ref{subsubsec:fpga-neurofuzzy}, follows the same partitioning: the ANFIS inference runs in programmable logic while the host core handles sensor I/O and grid synchronization.}
{Key challenges in SoC-based fuzzy systems include partitioning complexity, communication overhead, and power management. Partitioning is complicated by the fact that the fuzzy pipeline contains stages of very different computational weight. Fuzzification and rule evaluation are parallel and benefit from hardware, whereas defuzzification, with its division and normalization, may be cheaper in software if the host core is otherwise idle. The optimal split therefore depends on the duty cycle, the slack available on the host, and the cost of moving intermediate results across the AXI interconnect. Communication overhead is not negligible: every transfer between the processing system and the programmable logic consumes bus cycles, and AXI-Stream transfers require buffering that is itself a resource. A design that offloads the entire pipeline to hardware may save host cycles but pay for them in DMA setup and interrupt latency, whereas a design that keeps everything in software may avoid those costs but miss the real-time deadline. Power management adds a third dimension, because the programmable logic and the processing system can be placed in different low-power states, and transitions between them cost both energy and latency. The existing literature treats the partitioning decision as given rather than as a design variable, and no work provides a quantitative model for choosing the split. This absence is one of the gaps identified in Section~\ref{sec:discussion}.}

\subsection{ASIC and Custom VLSI Implementations}
\label{subsec:asic}

ASICs and custom VLSI designs target fixed-function, high-volume deployment at the opposite end of the flexibility-performance spectrum from FPGAs. While ASICs offer no post-fabrication reconfigurability, they deliver \textit{peak performance and energy efficiency} through custom datapaths, optimized floor planning, and process-specific circuit techniques. These implementations suit high-volume, power-sensitive, or safety-critical applications where the fuzzy algorithm is fixed and fully characterized prior to tape-out.
The cost structure of ASICs differs from that of FPGAs: high non-recurring engineering (NRE) costs (design, verification, mask fabrication) but low per-unit costs (silicon, packaging, testing). Consequently, ASIC fuzzy processors are economically justified for applications with production volumes exceeding thousands or millions of units, or where the {PPA} requirements cannot be met by programmable platforms.
Modern ASIC implementations use advanced process technologies (\textit{e.g.,} 130nm, 45nm CMOS) to achieve gigahertz-range operating frequencies and microwatt-level power consumption. The primary design considerations include custom datapath design, deep pipelining, power gating, and multi-bank memory hierarchies optimized for fuzzy inference.
Table~\ref{tab:asic_summary} provides a summary of representative ASIC and custom VLSI fuzzy implementations across all architecture categories. The following subsections discuss each category in detail.

\subsubsection{Digital ASIC Fuzzy Processors and Accelerators}
\label{subsubsec:asic-digital}

\begin{table*}[]
\centering
\caption{Summary of ASIC and custom VLSI fuzzy logic implementations}
\label{tab:asic_summary}
\renewcommand{\arraystretch}{1.3}
\setlength{\tabcolsep}{5pt}
\footnotesize
\begin{tabular}{@{}
    >{\centering\arraybackslash}m{0.6cm}
    >{\raggedright\arraybackslash}m{3.2cm}
    >{\raggedright\arraybackslash}m{2.6cm}
    >{\raggedright\arraybackslash}m{2cm}
    >{\centering\arraybackslash}m{0.8cm}
    >{\raggedright\arraybackslash}m{4cm}
    @{}}
\toprule
\rowcolor{headerblue}
{\color{white}\textbf{Ref.}} &
{\color{white}\textbf{Application}} &
{\color{white}\textbf{Architecture}} &
{\color{white}\textbf{Technology}} &
{\color{white}\textbf{Rules}} &
{\color{white}\textbf{Key Contribution}} \\
\midrule
\cite{talaska2019parallel} & Fuzzy logic networks & Asynchronous CMOS & 130nm CMOS & N.A & 2ns delay, 17nW power, fully asynchronous \\
\rowcolor{rowgray}
\cite{talaska2023novel} & Programmable fuzzy controller & Micro-programmed & 130nm CMOS & N.A & 200 MSamples/s, <1mW, AI-tunable \\
\cite{imamguluyev2025fuzzy} & NN accelerator optimization & Fuzzy-driven ASIC & CMOS & 34 & 15-20\% power reduction, 12-18\% speed increase \\
\rowcolor{rowgray}
\cite{jalilvand2020fuzzy} & Fuzzy inference engine & Unary computing & 45nm CMOS & 81 & 82\% area saving, 46\% power reduction \\
\cite{9969358} & Image noise reduction & Stochastic computing & 45nm CMOS & 16 & 91.8\% area saving, 84.7\% power reduction \\
\bottomrule
\multicolumn{6}{@{}p{15.2cm}@{}}{\footnotesize Note: \textbf{N.A} denotes \textit{Not Applicable} for entries where the concept of fuzzy rules is not defined.} \\
\end{tabular}
\vspace{-1.5em}
\end{table*}

Digital ASIC fuzzy processors employ custom datapaths, deep pipelining, and parallel rule evaluation to achieve superior throughput and energy efficiency compared to FPGA and software counterparts. Unlike FPGAs, which implement logic using LUTs and programmable interconnects with significant overhead, ASICs implement logic using standard cells and customized datapaths with minimal overhead, enabling higher operating frequencies (GHz range) and lower power consumption.

Talaska \cite{talaska2019parallel} presents parallel, asynchronous fuzzy logic systems realized in CMOS technology, exploiting asynchronous design to reduce power consumption and improve speed. The absence of a global clock in asynchronous designs eliminates clock distribution power and enables event-driven computation, where circuits only switch when inputs change. This approach is particularly attractive for fuzzy systems with sparse input activity. The circuits are designed in 130nm CMOS technology and operate fully asynchronously, with data processing rates at the nanosecond scale (approximately 2 ns delay) and power dissipation of approximately 17 nW for OR neurons and 15 nW for AND neurons. A prototype chip incorporating the full adders used in these operators has been verified through laboratory measurements.

\begin{figure*}[]
\centering
\includegraphics[
    width=1.6\columnwidth,
    trim=0.5cm 0cm 0.5cm 0cm,
    clip
]{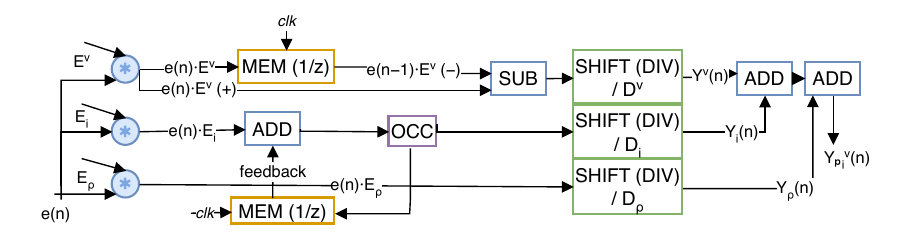}
\vspace{-1.em}
\caption{General structure of the parallel, asynchronous, hardware-programmable PID controller. The error sample $e(n)$ is multiplied in parallel by the fixed-point coefficients $E_D$, $E_I$, and $E_P$ (asterisk nodes, blue); in the derivative channel, a 1/z memory (orange, clocked) stores $e(n{-}1)\cdot E_D$ for the subtraction, while the integrating channel accumulates through an adder with an overflow-control component (OCC, purple) and its own delay memory. Each channel is normalized by an asynchronous shift-based divider ($/D_D$, $/D_I$, $/D_P$, green) before the two output adders combine $Y_D(n)$, $Y_I(n)$, and $Y_P(n)$ into $Y_{PID}(n)$; all datapath blocks operate combinationally without a global clock \cite{talaska2023novel}.}
\label{fig:talaska_pid}
\vspace{-1em}
\end{figure*}

Talaska \textit{et al.} \cite{talaska2023novel} propose a novel hardware-programmable PID controller adapted to cooperate with AI tuning algorithms for real-time adaptation. \textcolor{black}{The general structure of this parallel, asynchronous, hardware-programmable PID controller is illustrated in Fig.~\ref{fig:talaska_pid}, which details how the error sample is multiplied in parallel by fixed-point coefficients, processed through derivative and integrating channels with their respective delay memories and overflow-control components, and normalized by asynchronous shift-based dividers before final summation.} The controller uses a parallel and asynchronous structure where each of the P, I, and D blocks operates as a separate channel with its own multibit multiplier, summing circuit, and delay line. Designed in 130nm CMOS technology with 8-bit signal resolution, the circuit features a chip area below 0.1 mm\textsuperscript{2}, achieves data rates of 200-330 MSamples/s, and dissipates less than 1 mW. The micro-programmed architecture allows the rule base and membership function parameters to be updated under software control, offering a degree of flexibility rare in ASIC designs and bridging the gap between fixed-function ASIC efficiency and the adaptability required for modern AI-integrated systems.

Imamguluyev \textit{et al.} \cite{imamguluyev2025fuzzy} explore the use of fuzzy logic in ASIC optimization for neural network acceleration, linking fuzzy control to accelerator tuning. Using MATLAB's Fuzzy Toolbox with a Mamdani-based inference system and 34 fuzzy rules, the authors demonstrate that a fuzzy logic-based optimization engine can dynamically adjust neural network accelerator parameters (voltage, frequency, batch size) to achieve optimal energy-delay product. The simulation results show 15-20\% power consumption reduction and 12-18\% processing speed improvement compared to traditional deterministic optimization models.

Design considerations for digital ASIC fuzzy processors include custom instruction set architectures (ISAs) optimized for fuzzy operations, multi-bank memory hierarchies with separate banks for membership functions and rule storage, aggressive power gating techniques, and scalable architecture templates for different rule base sizes.

\subsubsection{Stochastic and Unary Fuzzy Architectures}
\label{subsubsec:asic-stochastic}

{Section~\ref{subsubsec:fpga-stochastic} discussed 
FPGA maps of unary and stochastic fuzzy engines. This subsection focuses on custom-silicon realizations, where bit-stream generation and counting are realized at the circuit level.}

{The unary engine of Jalilvand \textit{et al.}~\cite{jalilvand2020fuzzy, najafi2026method}, introduced in Section~\ref{subsubsec:fpga-stochastic} as an FPGA design, was mapped to a 45nm CMOS process; the ASIC realization preserves the relative area and power savings of the FPGA baseline, confirming that the algorithmic advantage of unary computing survives the move to custom silicon. Estiri \textit{et al.}~\cite{9969358} report the same behavior for their stochastic filter, targeting camera and sensor nodes with tight silicon budgets.}

{The application of stochastic computing to fuzzy neural networks was recently demonstrated by Farhadi \textit{et al.}~\cite{farhadi2026scfnns}, who implement an AND/OR inference architecture and a training circuit using only logic gates. Their design eliminates the range-adjustment step required by conventional stochastic computing, because all operations in a fuzzy neural network are bounded to the unit interval. On the Iris, Wine, and Breast Cancer datasets, their FPGA implementation reduces look-up table usage by 36\% to 53\% in training and by 19\% to 43\% in inference relative to floating-point and fixed-point baselines, while eliminating digital signal processors entirely. Training is accelerated by more than 97\%, and power consumption is reduced by up to 17\%. These results reinforce the observation that fuzzy arithmetic maps naturally onto stochastic logic, and they extend the approach from standalone inference engines to trainable neuro-fuzzy architectures.}

{Similar to FPGA fabrics, these ASIC maps mainly improve absolute area and power. The algorithmic limitations (convergence latency, stream overhead, and approximation error) remain those already noted in Section~\ref{subsubsec:fpga-stochastic}.}

\subsubsection{Analog and Mixed-Signal Fuzzy Circuits}
\label{subsubsec:asic-analog}

Analog and mixed-signal implementations exploit the physical behavior of
transistors to realize fuzzy operations. Membership functions are generated
by transconductance amplifiers operating in subthreshold or strong
inversion, and min and max operations are realized through current-mode or
voltage-mode circuits. {Current-controlled BiCMOS dividers and
current-mirror-based multipliers, for example, allow the defuzzification
step to be computed directly in the analog domain, without an explicit
analog-to-digital conversion~\cite{bothe2026fuzzyhardware}.} These designs
are attractive for sensor interfaces, where analog signals can be processed
without analog-to-digital conversion, reducing both latency and power.

{Early analog fuzzy chips established the feasibility of the approach.
Togai and Watanabe~\cite{togai1986fuzzy} demonstrated a 16-rule analog
fuzzy inference engine in 3\,$\mu$m CMOS, and
Yamakawa~\cite{yamakawa1993fuzzy} extended the approach to current-mode
membership-function generation in 2\,$\mu$m CMOS. Both designs operated in
continuous time, with response times limited only by the settling of the
analog signal path, and consumed power proportional to the bias current
rather than to the clock frequency. Neither design reported
programmability of the membership-function shape after fabrication, which
limited their use to fixed applications.}

{Recent work has revisited analog fuzzy circuits with modern process
nodes and programmable parameters. Sefraoui \textit{et
al.}~\cite{sefraoui2025programmable} present a programmable CMOS
Bell-shaped membership-function generator using a differential pair in
45\,nm technology, achieving 11\,$\mu$W with tunable center and width via
$V_{\mathrm{ref}}$ and transistor sizing. Their follow-up
work~\cite{sefraoui2026design} characterizes noise in Bell-shaped and
Sigmoid CMOS membership-function generators, analyzing flicker and thermal
noise. The Sigmoid design achieves a noise floor below
1\,$\mu$V/$\sqrt{\mathrm{Hz}}$ beyond 10\,kHz, with 93\% lower input
noise than the Bell-shaped circuit at 10\,kHz, while consuming
12\,$\mu$W.}

{The four analog designs differ in what they report. The two early
chips~\cite{togai1986fuzzy,yamakawa1993fuzzy} report process node and rule
count but not power; the two recent
generators~\cite{sefraoui2025programmable,sefraoui2026design} report power
and noise but not rule count, because they implement a single membership
function rather than a complete inference pipeline. No analog design among the reviewed works reports both a complete pipeline and a measured power figure on a modern process node. This gap is one reason the analog
designs do not appear in Table~\ref{tab:asic_summary}: the table's columns
(rule count, power, frequency) cannot be filled for any single analog
design without combining figures from two or more papers. The limitation
is partly an artifact of the search protocol: analog fuzzy circuits
published before 2000 or outside the three indexed databases were reached
only through snowballing, and additional designs may exist in venues not
covered here.}

\subsubsection{Low-Power and Approximate VLSI Fuzzy Architectures}
\label{subsubsec:asic-approx}

Approximate computing reduces power consumption and silicon area in fuzzy
hardware when small accuracy losses are acceptable. Control systems and
classification tasks tolerate modest output errors in exchange for energy
savings, and approximate fuzzy architectures intentionally relax precision
in arithmetic operations, data paths, or memory access to achieve those
savings.

{None of the reviewed ASIC designs quantifies how accuracy changes as approximation increases. The approximation techniques
that appear in the literature fall into four groups. The first truncates
multipliers and adders in the fuzzification and defuzzification datapaths.
The second reduces bit-width in less critical computations, typically from
16 to 8 bits in the aggregation stage. The third replaces centroid
defuzzification with approximate methods such as maximum membership or
weighted average with reduced precision, as discussed in
Section~\ref{sec:background}. The fourth operates circuits at reduced
supply voltages with error-correction techniques.}
{Each of these four techniques has been evaluated in isolation, and
none of the reviewed ASIC designs combines them. This absence has two
consequences. First, no work reports the combined effect of, for
example, bit-width reduction and voltage overscaling on the same design.
Second, no work compares approximation strategies on a common
benchmark, so the relative merit of the four techniques remains unknown.
The stochastic designs in Table~\ref{tab:asic_summary} approximate
arithmetic by construction, but they do not report accuracy against a
binary baseline in a form that permits comparison with the approximate
techniques above.}

{The challenge specific to fuzzy hardware is that the pipeline
contains both arithmetic operations, which tolerate approximation, and
comparison operations, which do not: a membership function evaluated with
a truncated multiplier degrades gracefully, whereas a min or max operation
evaluated with an approximate comparator can invert the firing order of
two rules and change the defuzzified output discontinuously. No
design quantifies this distinction, and it is one of the gaps discussed in
Section~\ref{sec:discussion}.}

\subsection{Embedded, IoT, and TinyML Fuzzy Systems}
\label{subsec:embedded}

Embedded fuzzy systems run on commercial-off-the-shelf (COTS)
microcontrollers, microprocessors, or specialized edge AI chips under
tight memory, processing, and power limits. Unlike FPGA or ASIC
implementations, which may require dedicated silicon or external
components, embedded designs rely on firmware and, in some cases, small
on-chip accelerators.
{Three categories must be distinguished within this platform class.
\emph{Embedded fuzzy inference} refers to a conventional fuzzy system with
hand-designed membership functions and rules, executed in software on a
microcontroller. No learning takes place; the rule base is fixed at design
time. \emph{TinyML} refers to a machine learning model, typically a neural
network, that has been trained offline and deployed on a microcontroller
or similarly constrained device. The model is learned from data, and
inference is the only on-device computation. \emph{Hardware-accelerated
machine learning} refers to a dedicated accelerator, such as an FPGA
overlay, an ASIC block, or a coprocessor, that executes the learned model
at higher throughput or lower energy than a general-purpose core. This
section covers all three, but the distinction matters because their cost
structures and design trade-offs differ: embedded fuzzy systems trade rule
count against memory, TinyML systems trade model capacity against peak
activation memory, and hardware-accelerated systems trade silicon area
against throughput.}

The constraints of embedded platforms are severe: memory (kilobytes to
megabytes), clock frequency (MHz range), and power budget (milliwatts to
microwatts) are orders of magnitude lower than FPGA or ASIC counterparts.
Consequently, embedded fuzzy systems require aggressive optimization at the
algorithm, data representation, and implementation levels. Key techniques
include fixed-point arithmetic (8-bit to 16-bit), sparse rule storage,
table-based evaluation (pre-computing membership function outputs in
ROM/flash), and event-driven execution (only performing inference when
inputs change significantly).

{On an FPGA or an ASIC the architecture itself is the primary
contribution. On an MCU, the main design question is how the four-stage
pipeline of Fig.~\ref{fig:fig1} maps onto a sequential instruction stream
and a few kilobytes of memory. Four realization strategies dominate, and
they span the code-size, latency, and generality trade-off. In
\emph{direct interpretation}, rules and membership functions are stored as
data structures and evaluated at run time by a generic routine. This is
the approach of the open-source embedded libraries (eFLL, fuzzylite) and
of model-based C code generation. It is the most flexible and the most
widely used, but it pays for genericity with per-rule branching. In
\emph{full look-up-table evaluation}, the input/output surface is
pre-computed in flash, so that latency becomes constant and minimal; table
size, however, grows as $Q^{n}$ for $n$ inputs quantized to $Q$ levels,
which confines the method to two- or three-input controllers. In
\emph{indexed rule evaluation}, the antecedent structure is flattened into
a static index table so that only rules capable of firing are visited.
Mart\'inez-Gonz\'alez \textit{et al.}~\cite{martinez2025combustion} show
that this technique alone removes most of the rule-evaluation overhead of
a textbook Mamdani implementation. Finally, \emph{fixed-point Q-format
arithmetic} (typically Q8.8 or Q16.16) can be executed on the single-cycle
multiply--accumulate (MAC) and single-instruction multiple-data (SIMD)
units of Cortex-M4 and M7 cores, in the spirit of the CMSIS-DSP library.
Orthogonal to these four strategies is event-driven inference, where the
engine is invoked only when an input crosses a significance threshold.
This approach converts periodic computation into sporadic execution and is the main energy-saving mechanism on battery-powered nodes.}

Table~\ref{tab:embedded_summary} provides a summary of representative
embedded, IoT, and TinyML fuzzy implementations. The following subsections
discuss each category in detail.

\subsubsection{Microcontroller-Based Fuzzy Systems}
\label{subsubsec:embedded-mcu}

MCUs are the most ubiquitous embedded computing platform, found in
billions of devices worldwide. MCU-based fuzzy systems must operate within
severe constraints: limited flash/SRAM (kilobytes), low clock frequencies
(tens to hundreds of MHz), and tight power budgets (milliwatts). Despite
these limitations, optimized fuzzy implementations can achieve
sub-millisecond inference latency for moderate rule bases.

\begin{table*}[]
\centering
\caption{{Summary of embedded, IoT, and TinyML fuzzy logic
implementations, organized by the metrics that bind on a microcontroller:
arithmetic format, memory footprint, and inference latency. Abbreviations:
EFLC = embedded fuzzy logic controller; AHB = Advanced High-Performance
Bus; TSK = Takagi--Sugeno--Kang; PSO = particle swarm optimization; WSN =
wireless sensor network; MF = membership function; NN = neural network;
PV = photovoltaic; N.R. = not reported.}}
\label{tab:embedded_summary}
\renewcommand{\arraystretch}{1.3}
\setlength{\tabcolsep}{3pt}
\footnotesize
\begin{tabular}{@{}
    >{\centering\arraybackslash}m{0.6cm}
    >{\raggedright\arraybackslash}m{2.7cm}
    >{\raggedright\arraybackslash}m{2.1cm}
    >{\centering\arraybackslash}m{0.55cm}
    >{\raggedright\arraybackslash}m{2.0cm}
    >{\raggedright\arraybackslash}m{1.7cm}
    >{\raggedright\arraybackslash}m{1.5cm}
    >{\raggedright\arraybackslash}m{1.5cm}
    >{\raggedright\arraybackslash}m{3.2cm}
    @{}}
\toprule
\rowcolor{headerblue}
{\color{white}\textbf{}} &
{\color{white}\textbf{Application}} &
{\color{white}\textbf{Platform}} &
{\color{white}\textbf{Rules}} &
{\color{white}\textbf{Arithmetic}} &
{\color{white}\textbf{Footprint}} &
{\color{white}\textbf{Latency}} &
{\color{white}\textbf{Power/Energy}} &
{\color{white}\textbf{Key Contribution}} \\
\midrule
\cite{jeong2023design} & Autonomous mobile robots & Cortex-M0 + AHB accelerator & 53 & Fixed-point (22b int, 10b frac) & N.R. & 4.56$\times$ vs.\ software & N.R. & Tightly-coupled EFLC accelerator; 2.459 max error \\
\rowcolor{rowgray}
\cite{martinez2025combustion} & Combustion (fuel/air ratio) & MCU (real-time loop) & N.R. & Fixed-point & Low program + RAM (reported qual.) & 20\,ms settling & N.R. & Pointer-array rule evaluation; Mamdani vs.\ TSK cost \\
\cite{sahin2025coupled} & Coupled-tank level control & STM32 Cortex-M7 & 25 & Floating-point & N.R. & Real-time (10\,ms loop) & N.R. & FLC vs.\ feedback-linearized PID on a modern M7 \\
\rowcolor{rowgray}
\cite{souza2025fuzzy} & Wheelchair control & Low-cost MCU & N.R. & Fixed-point & N.R. & N.R. & N.R. & Defuzzification iteration step vs.\ precision/time \\
\cite{lopezguede2026anfis} & IoT node duty-cycle / energy & ESP32 & 27 & Float, embedded C & $<$30\,KB RAM & $<$2.5\,ms & 31.1\% energy saving & On-device ANFIS inference; 10-day field trial \\
\rowcolor{rowgray}
\cite{karnavas2025adaptive} & DC micro-motor speed & STM32 (on-chip PSO) & N.R. & N.R. & N.R. & Real-time HIL & N.R. & IT2-FLC+PID tuned by PSO \emph{on the MCU}; 28.3\%/56.7\% faster settling \\
\cite{pau2019fuzzy} & WSN duty-cycle (smart home) & PIC24FJ256GB108 (COTS) & 9 & Fixed-point & N.R. & N.R. & Substantial node energy reduction & Fuzzy sleep-time control \\
\rowcolor{rowgray}
\cite{martinmoreno2025online} & Home energy-source management (PV + battery) & Arduino Nano 33 BLE Sense (Cortex-M4F) & 9 & N.R. & 45.8\,KB RAM total; 180\,B fuzzy model & N.R. (hourly decisions) & N.R. & On-device micro-genetic tuning of MFs with a fixed rule base; 27\% vs.\ 25.7\% cost saving \\
\cite{shaout2026ontarget} & Automotive deceleration command & STM32H723ZG Cortex-M7 & 27 & Floating-point (FPU) & 34.4\,KB flash, 26.0\,KB RAM (+14.3/+2.4\,KB with NN) & 166.8\,$\mu$s (177.3\,$\mu$s with NN) & N.R. & Same-core benchmark of a Mamdani FLC and an NN-tuned FLC; NN tuning adds 6.3\% time \\
\bottomrule
\multicolumn{9}{@{}p{16.5cm}@{}}{\footnotesize Note: Rows reflect microcontroller-class platforms only; FPGA- and ASIC-based works are reported in Tables~\ref{tab:fpga_summary} and~\ref{tab:asic_summary}.} \\
\end{tabular}%
\vspace{-1em}
\end{table*}

\begin{figure}[]
\centering
\includegraphics[
    width=\columnwidth,
    trim=0.5cm 0.5cm 0.5cm 0cm,
    clip
]{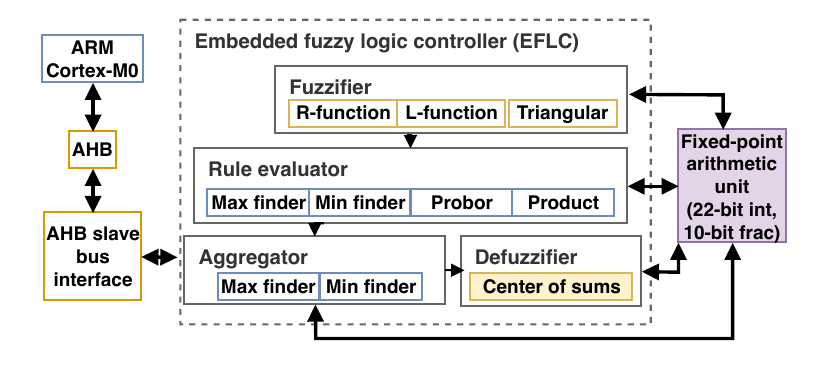}
\caption{Architecture of the EFLC integrated with the ARM Cortex-M0. The
processor (blue) accesses the EFLC (dashed region) through the AHB and its
slave bus interface (orange, thick bidirectional buses); inside the
accelerator, the fuzzifier (yellow: R-function, L-function, and triangular
membership units), the rule evaluator (blue: max finder, min finder,
probor, and product operators), the aggregator (blue: max/min finders),
and the center-of-sums defuzzifier (yellow) form the four-stage inference
pipeline. All stages share the fixed-point arithmetic unit (purple) with a
22-bit integer and 10-bit fractional representation, which enables the
reported 4.56$\times$ speedup over the software-only Cortex-M0
implementation~\cite{jeong2023design}.}
\label{fig:jeong_eflc}
\vspace{-1em}
\end{figure}

Jeong \textit{et al.}~\cite{jeong2023design} present an EFLC specifically
designed for autonomous mobile robots (AMRs) with an increasing number of
fuzzy rules. The architecture of this EFLC integrated with the ARM
Cortex-M0 is shown in Fig.~\ref{fig:jeong_eflc}, illustrating how the
processor accesses the EFLC through the AHB and its slave bus interface,
with the fuzzifier, rule evaluator, aggregator, and center-of-sums
defuzzifier forming the four-stage inference pipeline, all sharing a
fixed-point arithmetic unit with 22-bit integer and 10-bit fractional
representation. Experimental results demonstrate a maximum execution time
gain of 4.56x compared to software-only implementation, with a maximum
absolute error of 2.459 for output values, which is negligible for motor
voltage control applications.

In contrast to Jeong \textit{et al.}~\cite{jeong2023design}, which
offloads the pipeline to dedicated logic, a second body of work keeps the
engine in software and optimizes the cost algorithmically.
Mart\'inez-Gonz\'alez \textit{et al.}~\cite{martinez2025combustion}
control the fuel/air ratio of a premixed flame, reducing rule evaluation
to indexed accesses via a pointer-array scheme; the controller settles in
20\,ms at markedly lower computational cost than a textbook Mamdani
formulation, and the authors quantify the cost gap between Mamdani and TSK
inference on the same core. Sahin \textit{et al.}~\cite{sahin2025coupled}
benchmark a fuzzy controller for a coupled-tank system against a
feedback-linearized PID on an STM32 Cortex-M7, showing that an M7-class
core absorbs a 25-rule Mamdani inference within a 10\,ms control period
without any accelerator. Souza \textit{et al.}~\cite{souza2025fuzzy}
isolate defuzzification as the dominant cost on low-cost MCUs,
characterizing how the centroid iteration step trades precision against
processing time. At the IoT \emph{node} level, Pau and
Salerno~\cite{pau2019fuzzy} regulate sensor-node sleep intervals from
throughput, workload, and battery state with a nine-rule controller on a
COTS PIC24FJ, cutting energy with a footprint small enough to coexist with
the networking stack.

{Shaout and Castaneda-Trejo~\cite{shaout2026ontarget} report one of
the few on-target cost measurements in this group. They run a 27-rule
max--min Mamdani controller that computes automotive deceleration commands
on an STM32H723ZG (Cortex-M7) and compare it, on the same core, with a
variant in which a small neural network of 426 parameters shifts the
membership-function breakpoints. Timed with the core cycle counter, the
baseline needs 166.8\,$\mu$s per inference and the tuned version
177.3\,$\mu$s. The neural layer therefore adds 6.3\% to the execution
time, together with 14.3\,KB of flash and 2.4\,KB of RAM. A direct C
implementation of the same tuning policy runs 33.8$\times$ faster than the
network, so the authors argue that the learned layer is justified by
easier retraining and updates rather than by speed.}

{These works indicate that the binding constraint on an MCU is
rarely multiply-accumulate throughput. It is the memory traffic of rule
evaluation and the division required by defuzzification.
Table~\ref{tab:embedded_summary} shows the asymmetry directly: latency is
reported by seven of the nine works, memory footprint by four, and power
or energy by only two. The metrics that are easy to obtain from a
debugger are the ones that appear; the metrics that determine battery
life are the ones that are omitted.}

\subsubsection{TinyML and Edge-AI Fuzzy Systems}
\label{subsubsec:embedded-tinyml}

TinyML refers to deploying machine learning models on ultra-low-power
embedded devices, typically under 100\,KB of memory. Fuzzy inference, with
low compute cost and explicit rules, fits that budget. On the same
hardware, the main limit for MCU deployment is \emph{peak activation
memory}~\cite{lin2020mcunet}, not parameter count. Even aggressively
quantized networks remain expensive at this scale.

{A learned line-segment detector deployed on an STM32F746 with
320\,KB of SRAM requires a peak-activation footprint of approximately
256\,KB for a 25k-parameter int8 model, which leaves less than a quarter of
the device budget for the rest of the application. A fuzzy inference
engine, by contrast, costs two to three orders of magnitude less memory.
The ANFIS controller of Lopez-Guede~\cite{lopezguede2026anfis} executes in
under 2.5\,ms within 30\,KB of RAM on an ESP32, and the embedded
accelerator of Jeong~\cite{jeong2023design} keeps the entire pipeline in
dedicated logic with a 22-bit fixed-point datapath. Neither figure
approaches the peak-activation footprint of a quantized neural network of
comparable task complexity.}

{Fuzzy inference therefore fits the TinyML design space with a
kilobyte-scale memory footprint, data-independent worst-case latency, and
inherent interpretability. This suggests a complementary rather than
competitive relationship with neural networks: a quantized network can
perform feature extraction, while a compact fuzzy layer handles the
interpretable, safety-critical decision-making. Such partitioning is well
established in FPGA-based neuro-fuzzy
systems~\cite{lin2025implementation,hermassi2024zynq}. On a commodity
Cortex-M processor, only a first step has been reported: Shaout and
Castaneda-Trejo~\cite{shaout2026ontarget} run a small neural network beside
a Mamdani controller on a Cortex-M7, but the network only adjusts
membership-function breakpoints and does not extract features from raw
signals. A full pipeline of this kind has not yet been demonstrated on such
a part, and no fuzzy workload is currently represented in MLPerf
Tiny~\cite{banbury2021mlperftiny}.}

\subsubsection{Adaptive and On-Device Learning Fuzzy Systems}
\label{subsubsec:embedded-adaptive}

While most embedded fuzzy systems are fixed-function, with rule bases and
membership functions defined at design time, there is growing interest in
\emph{adaptive fuzzy systems} capable of learning and adjusting parameters
on-device. This capability is particularly valuable in applications where
environmental conditions change unpredictably and manual retuning is
impractical. On-device learning requires storage for gradients and updated
parameters, multipliers and accumulators for parameter updates, and careful
management to avoid interference with real-time inference.

Two approaches that the literature often groups together as on-device adaptation should be distinguished. The first is \emph{offline training with on-device
inference}. Lopez-Guede~\cite{lopezguede2026anfis} exemplifies this
approach: a first-order Takagi--Sugeno ANFIS with three Gaussian membership
functions per input (temperature, light, battery voltage) is trained
offline, translated to embedded C, and deployed on an ESP32 node that
adapts its deep-sleep interval to environmental conditions. A ten-day field
deployment achieved a 31.1\% energy reduction compared to a fixed-interval
baseline (RMSE = 9.6\,s), with inference completing in under 2.5\,ms within
30\,KB of RAM, a rare quantitative assessment of neuro-fuzzy inference cost
on a COTS microcontroller. The adaptation itself, however, is performed
offline and does not occur on the device.

The second approach, \emph{on-device parameter optimization}, remains rare.
Karnavas \textit{et al.}~\cite{karnavas2025adaptive} implement PSO directly
on an STM32 to tune an interval type-2 fuzzy PID controller for a DC
micro-motor, within a HIL framework that explicitly accounts for processing
delay and memory constraints. The resulting FT2-PID achieves 28.3\% and
56.7\% faster settling times than embedded-PSO-tuned PIDF and PI baselines.
{Mart\'in-Moreno \textit{et al.}~\cite{martinmoreno2025online} bring
this optimization to a TinyML device. Their Mamdani model, with two
inputs, one output, and nine expert rules, selects the energy source of a
home with solar panels and a battery, and runs on an Arduino Nano 33 BLE
Sense (Cortex-M4F). Once every 24\,h, a micro genetic algorithm ($\mu$CHC)
with ten chromosomes and 100 trials retunes the position and width of each
membership function on the device, using the buffered data of the previous
day. The rule base stays fixed, so the model remains interpretable. The
fuzzy model occupies 180\,B, and the whole program uses 45.8\,KB of RAM,
about 17\% of the device. On eleven months of recorded data, the adaptive
model raised the cost saving from 25.7\% for the initial static model to
27\%.} Two observations follow from these works. First, on-MCU adaptation is feasible
when the update rule is derivative-free and episodic, evaluated between
control cycles, rather than per-sample gradient-based, as the former
tolerates the absence of a floating-point unit (FPU) and limited RAM.
Second, interval type-2 inference, whose iterative type reduction is
particularly challenging for real-time hardware, becomes viable on an MCU
only when type reduction is replaced by a closed-form approximation.
Neither aspect has been systematically investigated, and no reviewed work
demonstrates true per-sample online learning of a fuzzy rule base on a
commodity microcontroller.


\section{Cross-Platform Comparison}
\label{sec:cross-platform}
{

The three platform classes are compared on the following criteria: latency,
power, flexibility, development cost, and rule-base size. Table~\ref{tab:platform_comparison} lists the entries
drawn from the works discussed in Section~\ref{sec:taxonomy}, which
makes it a decision aid rather than a ranking. The appropriate choice
depends on which criterion binds for a given application.
The \textit{latency} ranges of the three classes do not overlap. FPGA
and ASIC designs operate in the nanosecond to microsecond range. The
FPGA controller of Loukil~\textit{et al.}~\cite{loukil2020design}
executes a 25-rule Sugeno inference in 0.1127\,ms, and the
asynchronous ASIC neurons of Talaska~\cite{talaska2019parallel} settle
in approximately 2\,ns. Within the ASIC class, analog circuits operate
in continuous time, with response limited only by the settling of the
analog signal path. The early chips of Togai and
Watanabe~\cite{togai1986fuzzy} and Yamakawa~\cite{yamakawa1993fuzzy}
fall in this sub-class. MCU implementations operate in the microsecond
to millisecond range. The embedded accelerator of Jeong~\textit{et
al.}~\cite{jeong2023design} is 4.56 times faster than a software-only
Cortex-M0 implementation, and the ANFIS controller of
Lopez-Guede~\cite{lopezguede2026anfis} completes an inference in under
2.5\,ms on an ESP32.

\begin{table*}[]
\centering
\caption{{Comparison of the three hardware platform classes on the
criteria that determine platform selection. Entries report the range
observed across the works discussed in Section~\ref{sec:taxonomy}.
N.R.~=~not reported by the works in that class. Typical rule counts
reflect the majority of implementations discussed in each subsection;
individual designs outside these ranges exist. The ASIC row covers both
digital and analog realizations; the two sub-classes are distinguished
in the Latency, Power, and Flexibility columns where their ranges
differ.}}
\label{tab:platform_comparison}
\renewcommand{\arraystretch}{1.3}
\setlength{\tabcolsep}{5pt}
\footnotesize
\begin{tabular}{@{}
    >{\raggedright\arraybackslash}m{1cm}
    >{\centering\arraybackslash}m{1.5cm}
    >{\centering\arraybackslash}m{1.5cm}
    >{\raggedright\arraybackslash}m{5cm}
    >{\raggedright\arraybackslash}m{2.4cm}
    >{\centering\arraybackslash}m{1cm}
    >{\raggedright\arraybackslash}m{3.0cm}
    @{}}
\toprule
\rowcolor{headerblue}
{\color{white}\textbf{Platform}} &
{\color{white}\textbf{Latency}} &
{\color{white}\textbf{Power / Energy}} &
{\color{white}\textbf{Flexibility}} &
{\color{white}\textbf{Development Cost}} &
{\color{white}\textbf{Typical Rules}} &
{\color{white}\textbf{Representative Works}} \\
\midrule
FPGA &
ns--\textmu s &
mW--W &
High: reconfigurable fabric, partial reconfiguration, rule base changeable without new silicon &
Medium: NRE limited to synthesis and place-and-route; no mask cost &
10--100 &
Loukil~\cite{loukil2020design}, Swami~\cite{swami2022fpga}, Jalilvand~\cite{jalilvand2020fuzzy}, Matusi~\cite{matusi2022fpga} \\
\rowcolor{rowgray}
ASIC (digital and analog) &
sub-ns--ns (digital); continuous (analog) &
nW--mW (digital); sub-\textmu W (analog) &
None to low (digital): fixed datapath, micro-programmed variants allow parameter updates but not architectural change; low (analog): fixed topology, membership-function shape programmable in recent designs &
High: mask set, verification, and fabrication NRE; justified only at volume &
10--100 (digital); 16--64 (analog) &
Talaska~\cite{talaska2019parallel}, Jiang~\cite{jiang2026memristive}, Imamguluyev~\cite{imamguluyev2025fuzzy}, Togai~\cite{togai1986fuzzy}, Yamakawa~\cite{yamakawa1993fuzzy}, Sefraoui~\cite{sefraoui2025programmable} \\
MCU / TinyML &
\textmu s--ms &
\textmu W--mW &
Medium: firmware update; rule base and membership functions changeable without hardware change, within memory budget &
Low: COTS part, no NRE; development cost is software only &
$<$50 &
Jeong~\cite{jeong2023design}, Lopez-Guede~\cite{lopezguede2026anfis}, Shaout~\cite{shaout2026ontarget} \\
\bottomrule
\end{tabular}%
\vspace{-0.75em}
\end{table*}

The three orders of magnitude between the ASIC and MCU ranges explain
why closed-loop control with sub-microsecond deadlines
(Section~\ref{sec:introduction}) is confined to FPGA and ASIC designs. 
A motor drive that samples every microsecond cannot be
serviced by a Cortex-M0 running the pipeline in software, and the
accelerator of Jeong~\textit{et al.}~\cite{jeong2023design} addresses
exactly this constraint. The same separation explains why the TinyML
use cases of Section~\ref{subsubsec:embedded-tinyml} tolerate MCU
latency: their sampling periods are on the order of seconds, and a
millisecond inference falls well within the deadline.

Power consumption spans six orders of magnitude across the three
classes, from sub-microwatt analog circuits to watt-level FPGA designs.
The asynchronous ASIC neurons of Talaska~\cite{talaska2019parallel}
dissipate approximately 17\,nW for OR neurons and 15\,nW for AND
neurons. Within the ASIC class, the analog membership-function
generators of Sefraoui~\textit{et
al.}~\cite{sefraoui2025programmable} consume 11\,\textmu W, and their
Sigmoid design~\cite{sefraoui2026design} consumes 12\,\textmu W. MCU
implementations fall between these extremes and overlap the lower end
of both the FPGA and ASIC ranges. The ANFIS controller of
Lopez-Guede~\cite{lopezguede2026anfis} reduces node energy by 31.1\%
relative to a fixed-interval baseline, and the fuzzy sleep-time
controller of Pau and Salerno~\cite{pau2019fuzzy} cuts node energy with
a footprint small enough to coexist with the networking stack. FPGA
implementations report milliwatt to watt figures. The TS-ANFIS of
Hermassi~\textit{et al.}~\cite{hermassi2024zynq} consumes 0.215\,W.


Flexibility and development cost are inversely related across the
digital platform classes. FPGA offers the most flexibility at medium
non-recurring engineering cost. The fabric can be reconfigured for a
different rule base, a different membership-function shape, or a
different inference variant without new silicon, and partial
reconfiguration allows the change at run time
\cite{loukil2020design, abbes2022fuzzy}. ASIC offers the least
flexibility at high NRE. The datapath is fixed at tape-out, and the
micro-programmed variants of Talaska~\textit{et
al.}~\cite{talaska2023novel} allow parameter updates but not
architectural change. Within the ASIC class, analog circuits fall
outside this digital trade-off: their flexibility is limited by
topology rather than by configuration, and the rule base is not
programmable even where the membership-function shape is. The design is
process-specific and requires custom layout, which places analog at the
highest NRE among the ASIC sub-classes. MCU offers an intermediate
flexibility at the lowest NRE: the firmware can be updated in the
field, and the rule base and membership functions can be changed
without hardware change, within the memory budget.

The rule-count ranges in Table~\ref{tab:platform_comparison} describe
the works discussed in Section~\ref{sec:taxonomy}, not the platforms
themselves. The MCU range is the narrowest, and the cause is the memory
and latency constraints of Section~\ref{subsubsec:embedded-mcu}, not a
property of the platform. A rule base of 53 rules appears in the
accelerator of Jeong~\textit{et al.}~\cite{jeong2023design}, the
largest in the class. The FPGA and ASIC ranges extend to 81 rules in
the unary engine of Jalilvand~\textit{et
al.}~\cite{jalilvand2020fuzzy}, and the analog sub-class is limited by
the number of membership-function generators that fit on the die. The
rule-count column therefore supports the scalability gap of
Section~\ref{sec:discussion} rather than a platform ranking. No class
in the reviewed corpus reaches the several-hundred-rule scale that
applications with tens of input variables require, and the MCU class is
furthest from it.

A design that requires reconfigurability after deployment cannot use an
ASIC, regardless of its power advantage. A design that requires
sub-microwatt operation cannot use an FPGA, regardless of its
flexibility. A design whose rule base must be retuned in the field
cannot use an analog circuit. No single platform dominates because the
criteria in Table~\ref{tab:platform_comparison} are not commensurable,
and the appropriate choice is determined by which constraint the
application cannot relax. A field that cannot compare platforms on a
common basis cannot identify which of them limits progress, and the
absence of a benchmark suite is the first consequence.
}
\section{Research Gaps and Future Directions}
\label{sec:discussion}

\begin{figure*}[]
\centering
\includegraphics[
    width=1.2\columnwidth,
    trim=0.5cm 0cm 0.5cm 0.5cm,
    clip
]{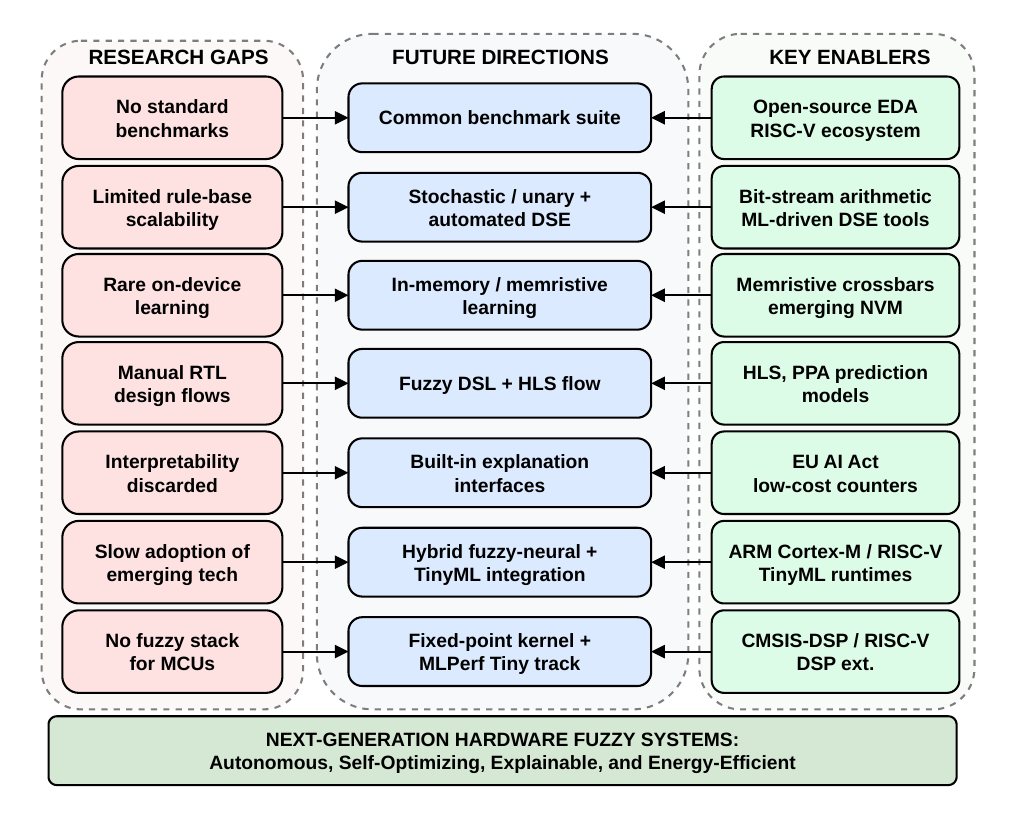}
\vspace{-1.75em}
\caption{Roadmap of the seven research gaps identified in this survey, the corresponding future research directions, and the enabling technologies that support them. The left entries are the gaps, middle entries are the near-term responses, and right entries are the enabling technologies; dashed arrows denote enabling relationships. The proposed directions build on early results in stochastic and unary computing \cite{jalilvand2020fuzzy,9969358}, memristive in-memory learning \cite{jiang2026memristive}, on-device neuro-fuzzy deployment \cite{lopezguede2026anfis,karnavas2025adaptive}, {on-device membership-function tuning and on-target cost measurement on microcontrollers \cite{martinmoreno2025online,shaout2026ontarget}}, and FPGA-based neuro-fuzzy co-design \cite{hermassi2024zynq,lin2025implementation}. {All directions converge toward hardware fuzzy systems with low energy use, on-device adaptation, and explainable rule-level outputs.}}
\label{fig:gaps_roadmap}
\vspace{-1em}
\end{figure*}

Despite decades of research and significant advances across FPGA, ASIC, and embedded platforms, several critical gaps remain unaddressed. These gaps span the entire design flow, from the absence of standardized evaluation frameworks and scalable architectures to the lack of automated design tools, on-device learning capabilities, and integration with emerging technologies. Fig.~\ref{fig:gaps_roadmap} maps each gap to a near-term response and its enabling technology.

\subsection{Standardized Benchmarks and Open-Source Reproducibility}

\textbf{Current Challenge:} The absence of a standardized benchmark suite for hardware fuzzy systems slows progress. Unlike the machine learning community, which benefits from established benchmarks such as ImageNet and MLPerf, fuzzy hardware researchers rely on disparate application-specific testbeds. {Most works omit key deployment metrics. Resource utilization is reported by only a minority of studies, operating frequency is often unspecified, and comprehensive power measurements are rare. This reporting gap makes cross-platform comparison difficult.}

\textbf{Future Direction:} {The field needs a common benchmark suite that covers} control problems (inverted pendulum, DC motor speed control), classification tasks (ECG anomaly detection), and signal processing benchmarks (noise filtering, image edge detection). {Standard} reporting guidelines should specify device family, operating frequency, resource utilization, power consumption, and inference accuracy for each benchmark. The open-source hardware movement{, exemplified by RISC-V, Chisel, and open-source EDA tools, offers} practical benefits for reproducibility, collaboration, and innovation. Useful components include a domain-specific language for fuzzy systems, an open-source compiler framework translating the language to platform-specific implementations, and open-source benchmark suites with reference implementations and reporting standards.

\subsection{Scalability and Rule Base Management}

\textbf{Current Challenge:} Scalability remains a challenge across all platforms. For FPGA and ASIC implementations, area scales roughly linearly with the number of rules, limiting practical implementations to a few hundred rules on mid-range devices. {The majority of FPGA implementations employ rule bases of a few dozen rules or fewer, with only a small fraction exceeding fifty rules.} The curse of dimensionality (exponential growth of rule combinations) presents major barriers to scaling for applications requiring tens or hundreds of input variables.

\textbf{Future Direction:} Stochastic and unary computing can cut rule-evaluation area at the cost of precision \cite{jalilvand2020fuzzy, 9969358, banitaba2025adversarial, jalilvand2022fast}. {Automated} design-space exploration tools, powered by machine learning and multi-objective optimization algorithms, could address this challenge. These tools would accept high-level requirements (target platform, performance goals, power budget, area constraints), explore the design space using fast approximate models, and generate Pareto-optimal designs that balance rule count, area, and accuracy.

\subsection{Online Learning and Adaptability}

\textbf{Current Challenge:} Hardware implementations with truly online, on-device learning remain rare. Only a few prior works address learning in hardware \cite{hermassi2024zynq, lin2025implementation, jiang2026memristive}, and even these implement learning primarily in software. {On microcontrollers, on-device tuning has so far been shown only for controller gains and membership functions, with the rule base kept fixed~\cite{karnavas2025adaptive, martinmoreno2025online}.} The barriers include computational demands (multipliers, dividers, gradient computation), memory bandwidth (frequent read-modify-write operations), and algorithmic challenges (standard learning algorithms are not inherently hardware-friendly).

\textbf{Future Direction:} Memristive crossbars for in-situ learning \cite{jiang2026memristive} reduce the cost of parameter updates relative to digital backprop on the same die. Hardware-aware update rules that respect memory and multiplier budgets are still needed. Longer term, accelerators that retune rules, clocks, or voltage at runtime from measured error or aging remain largely undemonstrated.

\subsection{Design Automation and Hardware-Aware Design}

\textbf{Current Challenge:} {Most prior works rely on manual register-transfer level (RTL) design, with only a small fraction adopting high-level synthesis or model-based design flows.} The primary challenges include the lack of domain-specific languages for fuzzy systems, platform-specific optimization requirements, and the need for automated design-space exploration tools. A related issue is the separation between algorithm design and hardware implementation: most fuzzy system design methodologies treat hardware constraints as afterthoughts, while hardware designers implement pre-existing algorithms without considering hardware-aware modifications.

\textbf{Future Direction:} Hardware-aware fuzzy system design would involve co-optimization of the fuzzy algorithm and its hardware implementation. For example, the choice of membership function shape should consider hardware cost; the number of rules should balance accuracy and resource usage; defuzzification methods should be selected based on hardware availability and precision. Tools for hardware-aware fuzzy design, including models that predict area, power, and latency for a given fuzzy configuration, would be valuable for guiding this co-optimization. Automated design-space exploration tools are the next step in this direction.

\subsection{Explainable Hardware AI}

\textbf{Current Challenge:} The demand for explainable artificial intelligence is growing rapidly, driven by regulatory requirements (e.g., the EU AI Act) and the need to audit decisions in safety-critical settings. {Fuzzy datapaths already compute rule firing strengths, yet most hardware designs export only the crisp output} and drop that intermediate evidence.

\textbf{Future Direction:} Hardware fuzzy systems should incorporate built-in diagnostic and explanation interfaces that report which rules were activated and their firing strengths, the contribution of each input variable to the final output, and the confidence or uncertainty associated with the output. Such explanation interfaces could be implemented with minimal additional hardware: counters and registers to track rule activation frequencies, and memory to store the top-k contributing rules. The integration of explainable hardware AI with TinyML is relevant here, as TinyML applications often operate in safety-critical contexts where explainability is required for regulatory compliance and user trust.

\subsection{Integration with Emerging Technologies and Hybrid Architectures}

\textbf{Current Challenge:} The fuzzy hardware research community has been slow to adopt emerging hardware technologies. Only a single work among the reviewed papers explores memristors \cite{jiang2026memristive}, and the adoption of open-source hardware design flows has been similarly limited. {Hybrid} fuzzy-neural architectures {, which combine the feature extraction capabilities of neural networks with the interpretability of fuzzy systems, remain} underexplored in hardware.

\textbf{Future Direction:} In-memory computing with memristors and other emerging memory technologies may improve energy efficiency by large margins. Hybrid fuzzy-neural accelerators that support both inference modes, with the ability to switch between them based on application requirements, remain an active research topic. The integration of fuzzy logic with TinyML ecosystems {, including standardized inference engines for ARM Cortex-M and RISC-V, model conversion tools, and benchmarking standards, will} extend the reach of fuzzy hardware to applications currently dominated by neural networks.

\subsection{Absence of a Fuzzy Software Stack for Commodity Microcontrollers}

\textbf{Current Challenge:} The microcontroller is the platform on which fuzzy logic should be most competitive, yet it is the one in which the community has invested least. Three deficits emerge from our review. First, no optimized fuzzy kernel library exists for Cortex-M or RISC-V comparable to {the neural network kernels provided by CMSIS-NN}. Every work discussed in Section~\ref{subsec:embedded} re-implements the inference pipeline from scratch, so that no shared fixed-point primitives exist and no basis for comparison is available. Second, with few exceptions, embedded fuzzy works report closed-loop quality such as rise time and overshoot, but omit the quantities that determine deployability, namely flash and RAM footprint, cycles per inference, and energy per inference. {Shaout and Castaneda-Trejo~\cite{shaout2026ontarget} show that such reporting is practical, giving flash, RAM, and cycle-counter timing measured on the target, yet even their study does not report energy per inference.} The density of \textbf{N.R.} entries in Table~\ref{tab:embedded_summary} makes this concrete. Third, no fuzzy workload appears in MLPerf Tiny~\cite{banbury2021mlperftiny}, leaving fuzzy inference invisible to the edge-AI ecosystem that is converging on the same hardware. A designer choosing between a quantized network and a fuzzy engine for a Cortex-M target therefore has no like-for-like data, even though the decisive criterion, peak activation memory~\cite{lin2020mcunet}, favours fuzzy systems.

\textbf{Future Direction:} A fuzzy counterpart of the TinyML software stack is needed. It would comprise an open-source fixed-point inference kernel for the Cortex-M and RISC-V DSP extensions, exposing Mamdani, Sugeno, and interval type-2 inference behind a single API with statically bounded memory; a reporting standard requiring footprint, cycles, and energy per inference alongside control metrics; and a fuzzy track within MLPerf Tiny. Because the worst-case latency of a fuzzy engine is data-independent, such a stack would also give safety-critical edge applications an interpretable inference engine with a provable worst-case execution time on a low-cost part.
{Among the seven research gaps identified, two should be addressed first}: (1) the establishment of standardized benchmarks and open-source reproducibility, as this first gap impedes progress across all other dimensions; and (2) the development of design automation tools for hardware-aware fuzzy system design, which would widen access to fuzzy hardware and accelerate innovation. The remaining gaps { (scalability, online learning, explainable AI, and emerging technology integration)}, while important, can be more effectively addressed once these foundational tools and standards are in place.

\section{Conclusion}
\label{sec:conclusion}

This presented a comprehensive review of hardware implementations
of fuzzy logic systems, organized through a platform-based taxonomy
that distinguishes FPGA-based designs, ASIC and custom VLSI realizations, and
embedded microcontroller and TinyML systems. This platform-centric
perspective provides a structured framework for understanding how the
underlying hardware substrate shapes architectural decisions, performance
outcomes, and design trade-offs.
{The study shows steady progress across all three platform
classes. FPGA-based implementations account for the largest share of the literature, driven by reconfigurability and rapid prototyping, with
recent work on neuro-fuzzy systems, type-2 fuzzy logic, and stochastic and
unary computing.}
{ASIC and custom VLSI implementations deliver peak performance and energy
efficiency. Prior designs report operating frequencies in the hundreds
of MHz and power consumption spanning the sub-milliwatt range down to tens of
nanowatts for asynchronous and stochastic designs. Work on memristive fuzzy
neural networks and in-memory computing may further improve energy
efficiency.}

{Embedded and TinyML systems support compact, low-power fuzzy inference. Prior implementations execute a complete inference pipeline within a
few hundred microseconds to a few milliseconds per inference, at memory
footprints of tens of kilobytes.}
{No single platform dominates across all metrics.} FPGAs offer
flexibility; ASICs deliver efficiency; analog circuits provide ultra-low
power; microcontrollers enable cost-effective IoT deployment; and TinyML
platforms balance power, performance, and cost for edge AI. Platform
selection depends on application-specific constraints including speed, power,
area, production volume, and development budget.

{Despite this progress, critical research gaps persist: the absence of
standardized benchmarks, scalability limitations for large rule bases,
immaturity of online learning in hardware, insufficient design automation,
and limited integration with emerging memory technologies and open-source
flows. Future advances will be driven by in-memory computing with emerging
technologies such as memristors, hybrid fuzzy-neural architectures,
on-device adaptation, explainable datapaths, and open design flows.
Standardized benchmarks and hardware-aware automation tools remain the most
actionable next steps.}

\bibliographystyle{IEEEtran}
\bibliography{References}


%
\vspace{-4em}
\begin{IEEEbiography}[{\includegraphics[width=1in,height=1.25in,clip,keepaspectratio]{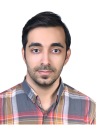}}]{Amir Hossein Jalilvand}
received his B.Sc. degree in Computer Engineering from Bu-Ali Sina University, Hamadan, Iran, and the M.Sc. degree in Computer Engineering - Computer Architecture from Iran University of Science and Technology, Tehran, Iran. He is currently pursuing his Ph.D. in Computer Engineering. His research interests include cellular networks, stochastic and unary computing, computer architecture, fuzzy logic, and machine learning. Mr. Jalilvand has authored several publications in these fields. He can be reached at jalilvand\_a@comp.iust.ac.ir.
\end{IEEEbiography}
\vspace{-4em}
\begin{IEEEbiography}[{\includegraphics[width=1in,height=1.25in,clip,keepaspectratio]{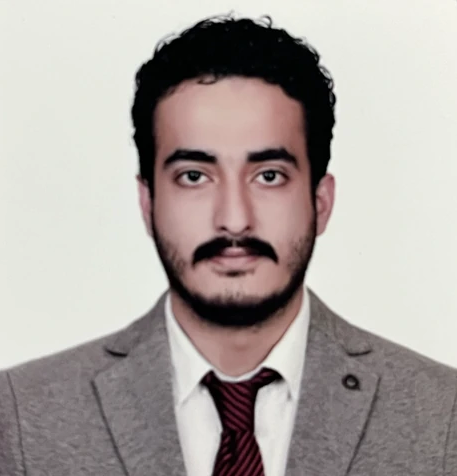}}]{Parsa Hassani Shariat Panahi}
received his B.Sc. degree in Computer Engineering from Azad University South Tehran Branch, Tehran, Iran, and the M.Sc. degree in Computer Engineering - Computer Networks from Iran University of Science and Technology, Tehran, Iran. His research interests include cellular networks, QoE assessment, telecommunication networks, wireless communication, and machine learning. He can be reached at parsa\_hassani@comp.iust.ac.ir.
\end{IEEEbiography}

\vspace{-4em}
\begin{IEEEbiography}[{\includegraphics[width=1in,height=1.25in,clip,keepaspectratio]{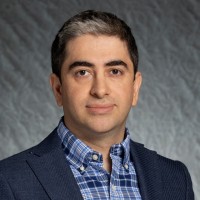}}]{M. Hassan Najafi}
received his Ph.D. in electrical and electronics engineering from the University of Minnesota-Twin Cities, Minneapolis, MN, USA, in 2018. He is currently an Associate Professor at the Electrical, Computer, and Systems Engineering Department at Case Western Reserve University. His research interests include stochastic and approximate computing, unary processing, in-memory computing, and hyperdimensional computing. He has authored/coauthored more than 110 peer-reviewed papers and has been granted 10 U.S. patents with more pending. Dr. Najafi received the NSF CAREER Award in 2024, the Best Paper Award at GLSVLSI'23 and ICCD'17, and the 2018 EDAA Outstanding Dissertation Award. Dr. Najafi is a senior member of IEEE and a senior member of NAI. He can be reached at najafi@case.edu.
\end{IEEEbiography}

\end{document}